\documentclass{aa}
\usepackage{natbib}
\bibpunct{(}{)}{;}{a}{}{,}
\usepackage{graphicx}
\usepackage{txfonts}
\usepackage[]{hyperref}
\hypersetup{
   colorlinks,%
   citecolor=magenta,%
   filecolor=blue,%
   linkcolor=magenta,%
   urlcolor=blue
}
\usepackage{float}
\usepackage{subcaption}
\usepackage{array}
\newcommand{\kB}{
k_{\mathrm{B}}
}

\newcommand{\derivative}[2]{
\frac{\mathrm{d}#1}{\mathrm{d}#2}
}

\newcommand{\indrm}[2]{
{#1_{\mathrm{#2}}}
}

\newcommand{\supcirc}[1]{
{\overset{\circ}{#1}}
}

\begin{document}

   \title{Dust evolution by chemisputtering during  protostellar formation}

   \author{Antonin Borderies
          \inst{1}
          \and
          Benoît Commerçon\inst{1}
          \and
          Bernard Bourdon\inst{2}
          }

   \institute{Centre de Recherche en Astrophysique de Lyon, ENS Lyon, UMR 5574 CNRS, Université de Lyon, 46 allée d’Italie, Lyon, France
         \and
         Laboratoire de Géologie de Lyon, ENS Lyon, UMR 5276 CNRS, Université de Lyon, 46 Allée d'Italie, Lyon, France
             }

   \date{}

% \abstract{}{}{}{}{} 
% 5 {} token are mandatory
 
  \abstract
  % context heading (optional)
  % {} leave it empty if necessary  
   {Dust grains play a crucial role in the modeling of protostellar formation, particularly through their opacity and interaction with the magnetic field. The destruction of dust grains in numerical simulations is currently modeled primarily by temperature-dependent functions. However, a dynamical approach could be necessary to accurately model the vaporization of dust grains.}
  % aims heading (mandatory)
   {We focused on modeling the evolution of dust grains during   star formation, specifically on the vaporization of the grains by chemisputtering. We also investigated the evolution of non-ideal magnetohydrodynamic resistivities and the Planck and Rosseland mean opacities influenced by the grain evolution.}
  % methods heading (mandatory)
   {We modeled the evolution of the dust by considering spherical grains at thermal equilibrium with the gas phase, composed only of one kind of material for each grain. We then took into account the exchange processes that can occur between the grains and the gas phase and that make the grain size evolve. We considered three materials for the grains: carbon, silicate, and aluminum oxide. Given a temporal evolution in temperature and density of the gas phase, we computed the evolution of a dust grain distribution. This evolution was then used to compute the non-ideal magnetohydrodynamic resistivities and the Planck and Rosseland mean opacities.}
  % results heading (mandatory)
   {We observed a significant dependence of the sublimation temperature of the carbon grains on the dynamical evolution of the gas phase. The application of our method to trajectories where the temperature and density of the gas decrease after the sublimation of a portion of the grain distribution highlights the limitations of current vaporization prescriptions in simulations.}
  % conclusions heading (optional), leave it empty if necessary 
   {The dynamical approach leads to more accurate results for the carbon grain quantity when the temperature and density of the gas evolve quickly. The dynamical approach application to collapse and disk evolution is then foreseen with its integration into hydrodynamic simulations.}


   \maketitle
%
%-------------------------------------------------------------------

%%%%%%%%%%%%%%%%%%%%%%%%%%%%%%%%%%%%%%%%%%%%%%%%%%%%%%%%%%%%%%%%%%%%%%%%%%%%%%%%%%%%%%%%%%%%
%%%%%%%%%%%%%%%%%%%%%%%%%%%%%%%%%%%%%%%%%%%%%%%%%%%%%%%%%%%%%%%%%%%%%%%%%%%%%%%%%%%%%%%%%%%%
\nolinenumbers{}
\section{Introduction}
\label{sec: introduction}

The formation of a protostar and its surrounding protoplanetary disk is a complex process that involves numerous physical and chemical phenomena. One of the key components of these processes is the evolution of dust grains. These grains are formed in the cold and dense regions of the interstellar medium (ISM). They represent approximately 1\% of the mass budget in the medium, and their size distribution is well modeled by a power law called the~\citet*{mathisSizeDistributionInterstellar1977} distribution (MRN). They play a crucial role in the formation of the protostar and the disk, including their blackbody  emission and opacity, which allow the cooling of the gas~\citep{semenovRosselandPlanckMean2003,omukaiThermalFragmentationProperties2005,tsuribeDustcoolinginducedFragmentationLowMetallicity2006}. They also serve as the main formation sites for $\mathrm{H_2}$ molecules, which heats up the medium~\citep{gouldInterstellarAbundanceHydrogen1963}. Furthermore, their coupling with the magnetic field regulates its evolution via non-ideal magnetohydrodynamic (MHD) effects~\citep{marchandChemicalSolverCompute2016,tsukamotoImpactDustSize2022}. The size of the grains, and notably the presence of small grains, is also a key parameter as it can significantly impact the formation and shape of the disk~\citep{zhaoProtostellarDiscFormation2016,tsukamotoCoevolutionDustGrains2023}.

It is fundamental to understand the evolution of dust grains during the protostellar formation to model accurately all the physical processes that are related to them, including the dust opacity and the non-ideal MHD resistivities. 
The main effects that can impact the evolution of  dust grains during the gravitational collapse of a molecular cloud are the interactions with the gas phase which remove (grain {vaporization or sublimation}) or add materials (grain {growth or condensation}), and notably through chemical reactions that occur on the surface of the grains (called {chemisputtering}), which leads ultimately to their total destruction when the temperature reaches a few thousand Kelvin. 
\citet{lenzuniDustEvaporationProtostellar1995} performed an analysis of these effects in the case of a protostellar core contraction; they modeled the vaporization of three  types of dust grains: carbonaceous, silicate, and aluminum oxide grains. The results of this study were then used to compute, among other things, the evolution of non-ideal MHD resistivities with temperature, and to build a resistivity table that has been  used in recent numerical simulations~\citep{marchandChemicalSolverCompute2016,vaytetProtostellarBirthAmbipolar2018}. Similarly, the modeling of the evolution of the dust opacity in numerical simulations is currently also made using an opacity table~\citep{vaytetSimulationsProtostellarCollapse2013}.

\begin{figure}[t]
   \centering
   \includegraphics[height = 6cm]{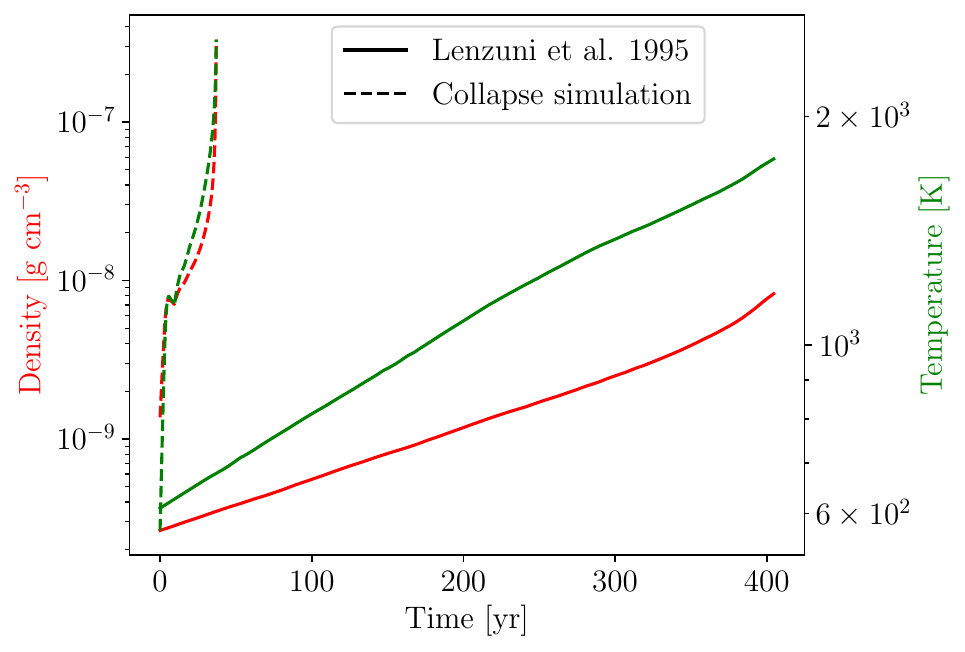}
   \caption{Comparison of the evolutions of temperature (in {green}) and density (in {red})  used by~\cite[{solid line}]{lenzuniDustEvaporationProtostellar1995}, with the evolution encounter in the central cell of a collapse simulation of a 1 solar mass cloud with the hydrodynamic code RAMSES ({dashed line}). The details of the simulation are given in Sect.~\ref{sec: dust_evolution_computation}.}
   \label{fig: lenzuni_vs_ramses}
\end{figure}

However, in this same study from~\citet{lenzuniDustEvaporationProtostellar1995}, the authors highlight the fact that the dust-gas interactions could have timescales similar to the protostar formation timescale (the free-fall time). Thus, the dust vaporization, the dust opacity, and the non-ideal MHD resistivities cannot be solely  functions of the temperature and density of the medium; instead,  a dynamic approach has to be taken. One limitation of the study from \citet{lenzuniDustEvaporationProtostellar1995} for the protostar formation is that they only considered a single evolution path for the temperature and the density of the medium. Since this evolution corresponds to a protostellar core contraction and not a gravitational collapse, their temperature and density increase much more slowly than  can be encountered in a collapse simulation, as shown in Fig.~\ref{fig: lenzuni_vs_ramses}. In young protostars, the continuous accretion and ejection and/or the rapid disk expansion can also cause  a rapid increase and decrease in temperature and density, which could lead to different sublimation limits~\citep[see][]{tsukamotoAshfallInducedMolecular2021, morbidelliFormationEvolutionProtoplanetary2024a}.

Thus, we propose to revisit the dust grain vaporization in the context of a protostar formation, with an approach similar to \citet{lenzuniDustEvaporationProtostellar1995}. In Sect.~\ref{sec: dust_evaporation_modelling} we present the model and the method used to compute the evolution of the dust grains. In Sect.~\ref{sec: dust_materials} we discuss the materials considered. In Sect.~\ref{sec: dust_evolution_computation} we present the results of the computation of the dust grain evolution in a protostellar formation context. Finally, in Sect.~\ref{sec: discussion} we discuss the implications of this study.

%%%%%%%%%%%%%%%%%%%%%%%%%%%%%%%%%%%%%%%%%%%%%%%%%%%%%%%%%%%%%%%%%%%%%%%%%%%%%%%%%%%%%%%%%%%%
%%%%%%%%%%%%%%%%%%%%%%%%%%%%%%%%%%%%%%%%%%%%%%%%%%%%%%%%%%%%%%%%%%%%%%%%%%%%%%%%%%%%%%%%%%%%
\section{Dust evolution modeling}
\label{sec: dust_evaporation_modelling}

To model the dust, we used the same set of assumptions as \citet{lenzuniDustEvaporationProtostellar1995}. A dust grain is defined as a collection of $N$ monomers (which can be a molecule or an atom) of volume~$\indrm{V}{m}$ and mass $\indrm{m}{m}$, defining the volume of the grain as $V = \indrm{V}{m} N$. We assumed that the grains are pure (i.e., each one is   composed of only one kind of monomer), and we also limit our study to spherical grains, defining their radius $a$ through the relation $V = \indrm{V}{m} N = \frac{4}{3}\pi a^3$, and their surface area corresponding to $\mathcal{A} = 4\pi a^2$.  Finally, we also supposed that the dust grains are always in thermal equilibrium with the gas phase, meaning that both share the same temperature $T$ at each time.

Given this set of assumptions, we could then model the evolution of the number of monomers in one dust grain following the method of \citet{gailPhysicsChemistryCircumstellar2013}. We start from the equation
\begin{equation}
   \derivative{N}{t} = \indrm{F}{gr} - \indrm{F}{vap},
\end{equation}
where $\indrm{F}{gr}$ and $\indrm{F}{vap}$ are respectively the rates of addition and subtraction of monomers on the grain through reaction with the gas phase. It is more convenient to express them in terms of reaction rate per unit surface area $J$, so that $F = \mathcal{A} J$. We can then use the relation between $N$ and $a$ to find the equation of evolution of the grain radius:
\begin{equation}
   \derivative{a}{t} = \indrm{V}{m} \left(\indrm{J}{gr} - \indrm{J}{vap} \right)~. 
   \label{eq: grain_radius_evolution}
\end{equation}

%%%%%%%%%%%%%%%%%%%%%%%%%%%%%%%%%%%%%%%%%%%%%%
\subsection{Grain growth}
\label{subsec: grain_growth}

There are several ways to add monomers to a grain. It notably depends on whether the monomer is stable in the gas phase or not. The presence of other chemical species in the gas phase can modify the reaction path and then the growth rate.

We start with the simplest reaction that can add a monomer to the grain, which is
\begin{equation}
   {\mathrm{M}_N}_{\mathrm{(s)}} + {\mathrm{M}_1}_{\mathrm{(g)}} \longrightarrow {\mathrm{M}_{N+1}}_{\mathrm{(s)}}~, 
   \label{eq: homomolecular_grain_growth}
\end{equation} 
where ${\mathrm{M}_N}_{\mathrm{(s)}}$ represents the grain with $N$ monomers, and ${\mathrm{M}_1}_{\mathrm{(g)}}$ represents the monomers in the gas phase. The reaction rate of this process can be expressed as
\begin{equation}
   \indrm{F}{gr} = \alpha \indrm{f}{col},
\end{equation}
where $\indrm{f}{col}$ is the collision frequency between a monomer and the grain, and $\alpha$ is the sticking coefficient, corresponding to the probability that a collision results in an absorption. The collision frequency in the case of a spherical grain can be expressed as
\begin{equation}
   \indrm{f}{col} = \frac{1}{4} \mathcal{A}  \indrm{v}{rel} \indrm{n}{m} = 4 \pi a^2 \sqrt{\frac{\kB T}{2\pi\indrm{m}{m}}} \indrm{n}{m}~,
\end{equation}
where $\indrm{v}{rel}$ is the mean relative thermal velocity between the dust grain and the monomers, which can be expressed as $\sqrt{\frac{8\kB T}{\pi\indrm{m}{m}}}$\footnote{We assume here that the grain is always perfectly coupled to the gas, such that there is no drift velocity between the grain and the gas phase.}, and $\indrm{n}{m}$ is the number density of monomers in the gas phase.
We then find the reaction rate per unit surface area as
\begin{equation}
   \indrm{J}{gr} = \alpha \sqrt{\frac{\kB T}{2\pi\indrm{m}{m}}} \indrm{n}{m}. 
\end{equation}

We can then extend this case to a more general reaction that can add a monomer to the grain
\begin{equation}
   {\mathrm{M}_N}_{\mathrm{(s)}} + \sum_{k=1}^{\indrm{N}{r}} \nu_k \mathrm{A}_k \longrightarrow {\mathrm{M}_{N+1}}_{\mathrm{(s)}} + \sum_{k=1}^{\indrm{N}{p}} \mu_k \mathrm{B}_k ,
   \label{eq: heteromolecular_grain_growth}
\end{equation}
where the $\nu$ and $\mu$ parameters are the stoichiometric coefficients of the reaction and the $\mathrm{A}_k$ and $\mathrm{B}_k$ are the reactants and products, respectively.
If this reaction has a limiting step in its kinetic path, we can estimate the growth rate by looking at the key reactant associated with this step. This reactant can be found through experimental results, but it is possible to get an idea by looking at the number densities $n_k$ of each reactant and comparing the different $n_k/\indrm{\nu}{r,}_k$ ratios. If one is significantly smaller than the others, the species associated with it is a good candidate to be the key reactant.
In this case, the growth rate per unit surface can be written similarly to our simple case and can be expressed as
\begin{equation}
   \indrm{J}{gr} = \alpha \sqrt{\frac{\kB T}{2\pi\indrm{m}{key}}} \frac{\indrm{n}{key}}{\indrm{\nu}{key}},
\end{equation}  
where $\indrm{n}{key}$, $\indrm{m}{key}$, and $\indrm{\nu}{key}$ are respectively the number density of the key reactant, its mass, and its associated stoichiometric coefficient.

%%%%%%%%%%%%%%%%%%%%%%%%%%%%%%%%%%%%%%%%%%%%%%
\subsection{Grain vaporization}
\label{subsec: grain_vaporization}

The removal of monomers from the grain can be caused either by ejection of a monomer from the surface of the grain by thermal agitation (thermal vaporization or free vaporization) or by a reaction with the gas phase (chemisputtering). In both cases these processes can be described as the reverse of reactions~\eqref{eq: homomolecular_grain_growth} and~\eqref{eq: heteromolecular_grain_growth}. A general reaction describing the evolution of the number of monomers in the grain can be written as
\begin{equation}
   {\mathrm{M}_N}_{\mathrm{(s)}} + \sum_{k=1}^{\indrm{N}{r}} \nu_k \mathrm{A}_k \rightleftharpoons  {\mathrm{M}_{N+1}}_{\mathrm{(s)}} + \sum_{k=1}^{\indrm{N}{p}} \mu_k \mathrm{B}_k .
   \label{eq: grain_evolution_reaction}
\end{equation} 
The presence of species $\mathrm{B}_k$ on the right-hand side indicates whether this is free vaporization (no $\mathrm{B}_k$) or chemisputtering (at least one $\mathrm{B}_k$). There are two ways to estimate the rate of vaporization. We can use the {detailed balance principle}, but in the case of chemisputtering, we can also use a similar approach to the growth model with a key species that is responsible for the chemisputtering. The second approach is more accurate, but requires experimental data to determine the reaction probability.

\subsubsection{Detailed balance principle}
\label{subsubsec: detailed_balance_principle}
This principle states that at equilibrium, the growth rate and the vaporization rate are equal for each reaction occurring at the surface of the grain. This principle can be expressed in terms of rate per unit surface area as
\begin{equation}
   \indrm{J}{vap} = \indrm{J}{gr,~eq} = \alpha \sqrt{\frac{\kB T}{2\pi\indrm{m}{key}}} \frac{\indrm{\supcirc{n}}{key}}{\indrm{\nu}{key}},
   \label{eq: detailed_balance_principle}
\end{equation}
where $\indrm{\supcirc{n}}{key}$ is the number density of the key reactant { (defined in Sect.~\ref{subsec: grain_growth})} when { the thermodynamical} equilibrium is reached. 
The total rate of evolution of the grain can then be written as
\begin{equation}
   \begin{aligned}
   \indrm{J}{tot} = \indrm{J}{gr} - \indrm{J}{vap} &= \alpha \sqrt{\frac{\kB T}{2\pi\indrm{m}{key}}} \frac{\indrm{n}{key}}{\indrm{\nu}{key}} \left(1 - \frac{\indrm{\supcirc{n}}{key}}{\indrm{n}{key}}\right) \\
     &= \alpha \frac{\indrm{p}{key}}{\indrm{\nu}{key}\sqrt{2\pi\indrm{m}{key}\kB T}}
     \left(1 - \frac{\indrm{\supcirc{p}}{key}}{\indrm{p}{key}}\right),
   \end{aligned}
   \label{eq: total_rate}
\end{equation}
where $\indrm{p}{key} = \indrm{n}{key}\kB T$ is the partial pressure of the key reactant.

All we need now is a method for estimating the number density and pressure at equilibrium of the key reactant. 
We  assume that we have a state out of equilibrium, with the partial pressure of the key species  denoted $\indrm{p}{key}$ and the partial pressure of all the reactants and products denoted $p_{\mathrm{A}_k}$ (with $k\neq \mathrm{key}$) and $p_{\mathrm{B}_k}$. We denote by $\indrm{\supcirc{p}}{key}$ the partial pressure of the key reactant that is needed to reach equilibrium, given the partial pressure of all the other reactants and products.
To compute $\indrm{\supcirc{p}}{key}$, we use thermochemical considerations: at equilibrium between the gas phase and the grain, reaction~\eqref{eq: grain_evolution_reaction} obeys the law of mass action which states that we have the following relation between the different partial pressures of the different species:
\begin{equation}
   \prod_{k=1}^{\indrm{N}{p}} p_{\mathrm{B}_k}^{\mu_k} = \indrm{\supcirc{p}}{key}^{\nu_{\mathrm{key}}} e^{-\indrm{\Delta}{r} G/RT} \prod_{k=1,~k\neq \mathrm{key}}^{\indrm{N}{r}} p_{\mathrm{A}_k}^{\nu_k} .
   \label{eq: law_of_mass_action}
\end{equation}
Here $\indrm{\Delta}{r} G$ is the Gibbs energy (or free enthalpy) per mole of reaction~\eqref{eq: grain_evolution_reaction}, and can be expressed as
\begin{equation}
   \indrm{\Delta}{r} G = \indrm{\Delta}{f} G(\mathrm{M}_{1 \mathrm{(s)}}) + \sum_{k=1}^{\indrm{N}{p}} \mu_k \indrm{\Delta}{f} G(\mathrm{B}_k) - \sum_{k=1}^{\indrm{N}{r}} \nu_k \indrm{\Delta}{f} G(\mathrm{A}_k),
\end{equation}
where   $\indrm{\Delta}{f} G$ is the Gibbs energy of formation per mole and $\mathrm{M}_{1 \mathrm{(s)}}$ represents a monomer on the dust grain. 

From the relation~\eqref{eq: law_of_mass_action}, we just need to isolate $\indrm{\supcirc{p}}{key}$ and divide by the actual partial pressure of the key species $\indrm{p}{key}$ to obtain

\begin{equation}
   \frac{\indrm{\supcirc{p}}{key}}{\indrm{p}{key}} = \left(\frac{\prod_{k=1}^{\indrm{N}{p}} p_{\mathrm{B}_k}^{\mu_k}}{ e^{-\indrm{\Delta}{r} G/RT} \prod_{k=1}^{\indrm{N}{r}} p_{\mathrm{A}_k}^{\nu_k}}\right)^{1/\indrm{\nu}{key}},
   \label{eq: pseudo_activity_eq}
\end{equation}
which gives us the ratio which appears in Eq.~\eqref{eq: total_rate}.
This ratio can be interpreted as the inverse of the {pseudoactivity} of the monomer on the grain~\citep{gailPhysicsChemistryCircumstellar2013}, which is often also called the {supersaturation ratio}. Thus, just by knowing all the partial pressures of the different species involved in the reaction, Eq.~\eqref{eq: pseudo_activity_eq} allows  us to compute the total rate of evolution of the grain~\eqref{eq: total_rate}.

\subsubsection{Kinetic approach}
\label{subsubsec: detailed_balance_principle}

The detailed balance principle approach of the vaporization works well when the current state is not far from  equilibrium. In situations where vaporization completely dominates the growth in reaction~\eqref{eq: grain_evolution_reaction}, the thermodynamic approach is limited as it does not take into account kinetic considerations for the vaporization of the grain (the $\alpha$ parameter is defined for the condensation kinetic process) that can reduce the effective vaporization rate, notably in the case  of chemisputtering. In this situation we model the vaporization rate per unit surface area as \citet{lenzuniDustEvaporationProtostellar1995} did, with the expression
\begin{equation}
   \indrm{J}{vap} = \indrm{Y}{key}\sqrt{\frac{\kB T}{2\pi\indrm{m}{key, vap}}} \frac{\indrm{n}{key, vap}}{\indrm{\mu}{key, vap}},
   \label{eq: evaporation_rate_lenzuni}
\end{equation}
where $\indrm{n}{key, vap}$ and $\indrm{\mu}{key, vap}$ are associated with the key species involved in the vaporization process, and the $\indrm{Y}{key}$ is the yield coefficient (or the reaction probability, similar to the sticking coefficient) of the key species, which needs to be determined experimentally. This expression is very similar to   expression~\eqref{eq: detailed_balance_principle}, but instead focuses on the presence of   species $\mathrm{B_{key}}$, which  is responsible for the chemisputtering of the grain. This approach is more accurate than the detailed balance principle, but we need to have experimental data to determine the yield coefficient. 

For this work, we used Eq.~\eqref{eq: evaporation_rate_lenzuni} to compute the total rate of vaporization of the grain if experimental data were available, otherwise we used the detailed balance principle~\eqref{eq: detailed_balance_principle}.

%%%%%%%%%%%%%%%%%%%%%%%%%%%%%%%%%%%%%%%%%%%%%%
\subsection{Numerical method}
\label{subsec: numerical_method}

\begin{figure}[t]
   \centering
   \includegraphics[height = 10cm]{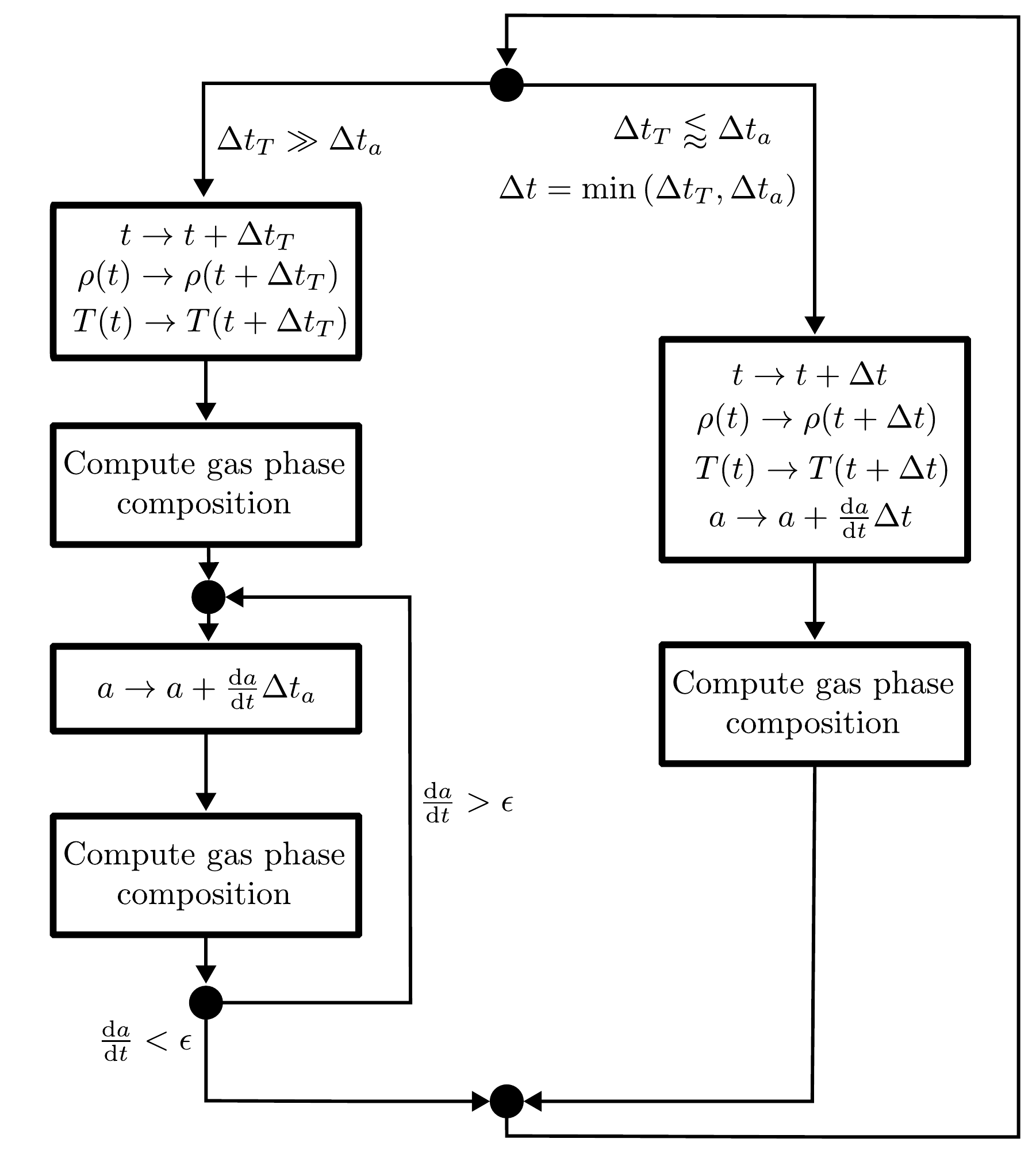}
   \caption{Schematization of the numerical method used to compute the evolution of a grain radius. 
            { When the radius $a$ evolves, we use Eq.~\eqref{eq: grain_radius_evolution} to compute the radius time derivative. The computation of the gas composition is done with FASTCHEM2~\citep{stockFASTCHEM2Improved2022}.}
            }
   \label{fig: numerical_method}
\end{figure}

Now that we have solved Eq.~\eqref{eq: grain_radius_evolution}, which allows us to compute the evolution of the grain radius, we need to compute this evolution for a given evolution of temperature $T(t)$ and density $\rho(t)$ for the gas phase surrounding the grain. As stated before, we make the assumption that the grain is always at thermal equilibrium with the gas phase. The numerical method is summarized in Fig.~\ref{fig: numerical_method}.

First of all, the computation of the radius evolution rate $J_{\mathrm{tot}}$ needs as input the composition of the gas phase, or at least the quantity of each species involved in the reactions with the grain. To this end, we use the code FASTCHEM2~\citep{stockFASTCHEM2Improved2022}, which  allows us to compute the chemical composition of the gas phase at equilibrium, given a temperature, pressure, and the relative abundance of each chemical element. This means that we make the assumption that the gas phase is always at chemical equilibrium, which can lead to some inaccuracies (see Sect.~\ref{sec: discussion} for more details).

For this calculation, the only requirement is to set the abundance of each element in the gas phase at the beginning of the computation, allowing us to determine the composition of the gas phase with FASTCHEM2. For this we need to set a global elemental abundance for the gas--dust mixture, and set the initial dust quantity. Then, we just have to remove the elements composing the dust from the global elemental abundance to get the composition of the gas phase.
Then, to integrate Eq.~\eqref{eq: grain_radius_evolution}, we need an integration scheme. We  use the Euler method, but better methods such as Runge-Kutta could be used.

The choice of the integration time step $\Delta t$ is crucial. To do this we need to consider the different timescales that exist in our problem: the temperature evolution timescale $\Delta t_T$, and the radius evolution timescale $\Delta t_a$. The density evolution timescale $\tau_\rho$ has the same order of magnitude as $\Delta t_T$ for the different gas evolutions we consider in this work, and is therefore not considered. To evaluate these  timescales, we need to define a maximum temperature step $\Delta T$ and a maximum radius step $\Delta a$. It is then possible to define the timescales as
\begin{equation}
   \begin{aligned}
      \Delta t_T &= \frac{\Delta T}{\derivative{T}{t}} \\
      \Delta t_a &= \frac{\Delta a}{\derivative{a}{t}} = \frac{\Delta a}{ \indrm{V}{m} J_{\mathrm{tot}}}.
   \end{aligned}
   \label{eq: time_scales}
\end{equation}
Several cases can occur:
\begin{enumerate}
   \item {$\Delta t_T \lessapprox \Delta t_a$: The grain evolves at the same speed as or more slowly than the temperature. The time step is then set as the smallest value between the two timescales, and we perform a single integration step.}
   \item $\Delta t_T \gg \Delta t_a$: The evolution of the grain is very fast compared to the evolution of the temperature; thus, we can consider that it reaches its equilibrium at a constant temperature. As we want to limit computation time, we first increase the time by $\Delta t_T$ to change the temperature and density, such that the grain is set out of equilibrium. We then set $\Delta t = \Delta t_a$, or a fraction of it, and perform multiple steps until the grain radius reaches its equilibrium with the gas phase $\indrm{a}{eq}(T)$. To know when the equilibrium is reached, we compare the derivative of the radius with a parameter $\epsilon$, and we stop the integration when the derivative is smaller than $\epsilon$.
\end{enumerate}

If the grain radius changes during the time step, it also means that it has modified the composition of the gas phase by removing or adding the materials composing the monomer. The number of monomers in the dust grains that change during the time step $\Delta t$ can be computed from the radius variation and is equal to
\begin{equation}
   \Delta N = \frac{1}{\indrm{V}{m}}\frac{4}{3}\pi\left(a(t+\Delta t)^3-a(t)^3\right).
\end{equation}
From this we can remove or add from the gas phase the chemical elements forming the monomers, and then recompute the new chemical composition of the gas with FASTCHEM2.

It should be  noted that the right-hand side of Eq.~\eqref{eq: grain_radius_evolution} does not depend on the radius of the grain. It is then possible to   evolve multiple grains with different radii at the same time, such that we can mimic the evolution of a grain size distribution.

%%%%%%%%%%%%%%%%%%%%%%%%%%%%%%%%%%%%%%%%%%%%%%%%%%%%%%%%%%%%%%%%%%%%%%%%%%%%%%%%%%%%%%%%%%%%
%%%%%%%%%%%%%%%%%%%%%%%%%%%%%%%%%%%%%%%%%%%%%%%%%%%%%%%%%%%%%%%%%%%%%%%%%%%%%%%%%%%%%%%%%%%%
\section{Dust materials}
\label{sec: dust_materials}

\begin{table}[b]
   \begin{tabular}{c|ccc}
      \hline
      Monomer & $\indrm{V}{m}$ (cm$^3$) & $\indrm{\rho}{grain}$ (g~cm$^{-3}$) & $\indrm{M}{d}/\indrm{M}{g}$ \\
      \hline
      C     &  $9.01\times 10^{-23}$ &  2.23 &  $2.35\times 10^{-3}$ \\
      $\mathrm{Mg_2SiO_4}$    &  $7.33\times 10^{-23}$ &  3.21 &  $2.05\times 10^{-3}$ \\
      $\mathrm{Al_2O_3}$    &  $4.24\times 10^{-23}$ &  4.02 &  $1.05\times 10^{-4}$ \\
      \hline

   \end{tabular}
   \caption{For each type of material, the values of the volume taken by each monomer in the grain $\indrm{V}{m}$, the density of the grain $\indrm{\rho}{grain}$, and the dust-to-gas ratio $\indrm{M}{d}/\indrm{M}{g}$ in the ISM for a solar composition.}
   \label{tab: dust_materials}
\end{table}

To apply this model to a protostellar formation problem, we need to set the composition of the dust that is present during the process. We choose a solar composition for the chemical elemental abundances, such that the three main species of grain that are present are carbonaceous grains, silicates (specifically olivine here), and aluminum oxides. 
{For the initial dust quantity, we follow~\citet{lenzuniDustEvaporationProtostellar1995}, assuming that most of the Si, Mg, and Al are locked in the grains, such that their elemental abundances set the dust quantities of silicates and aluminum oxides, while for the carbonaceous grains, it is 70\% of the carbon is locked in the dust.} 
Some of their properties are summarized in Table~\ref{tab: dust_materials}. For all the reactions that appear in this section, the values for the Gibbs energies of reaction are taken from~\citet{chaseNISTJANAFThermochemicalTables1998} and are given in Appendix~\ref{app: thermochemical_data}. We describe below how each of these species interacts with the gas phase.

%%%%%%%%%%%%%%%%%%%%%%%%%%%%%%%%%%%%%%%%%%%%%%
\subsection{Carbon}
\label{subsec: carbon}

\begin{figure}[t]
   \centering
   \includegraphics[height = 6.cm]{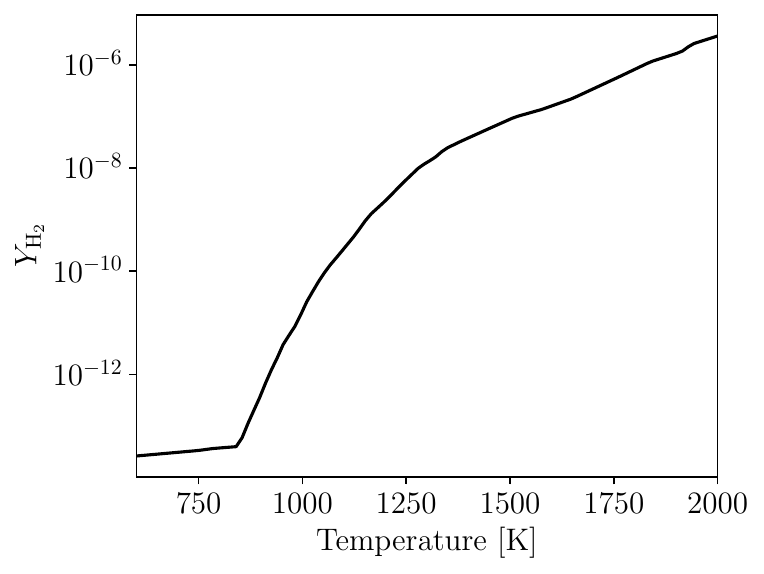}
   \caption{Yield coefficient of the chemisputtering by molecular hydrogen as a function of the temperature from the survey of~\citet{krakowskiSURVEYLITERATURECARBONHYDROGEN1970}.}
   \label{fig: yield_hydrogen}
\end{figure}

The monomers composing the carbonaceous grains are mainly carbon atoms. The first reaction with the gas phase that can then occur at the surface of this kind of grain is the adsorption of carbon atoms alone or in small groups of $i$ atoms (called $i$-mers, with $i$ typically equal to 2 or 3) in the gas phase, and their subsequent vaporization. This corresponds to the reaction
\begin{equation}
   {\mathrm{C}_{N}}_{\mathrm{(s)}} + \frac{1}{i}{\mathrm{C}_{i}}_{\mathrm{(g)}} \rightleftharpoons {\mathrm{C}_{N+1}}_{\mathrm{(s)}}~.
   \label{eq: free_vaporization_carbon}
\end{equation}
This reaction can be treated with the thermodynamic approach described in Sect.~\ref{subsubsec: detailed_balance_principle}, but it is also possible to use a kinetic approach, as described by~\citet{grassiDetailedFrameworkIncorporate2017} or~\citet{stahlerEvolutionProtostarsIII1981}, who considered the probability that a monomer can escape the grain using the Debye frequency, which takes into account the structure of the grain. Our calculations show that both approaches are equivalent, so we keep the thermodynamic approach for this study. Finally, the sticking coefficient for this reaction is about $0.3$ as shown in Table 3 of~\citet{lenzuniDustEvaporationProtostellar1995}.

Carbonaceous grains can also be subject to chemisputtering by three chemical species in the gas phase: atomic hydrogen, molecular hydrogen, and water molecules. The chemisputtering by atomic hydrogen takes the global form 
\begin{equation}
   {\mathrm{C}_{N+m}}_{\mathrm{(s)}} + n{\mathrm{H}}_{\mathrm{(g)}} \longrightarrow  {\mathrm{C}_{N}}_{\mathrm{(s)}} + {\mathrm{C}_{m}}{\mathrm{H}_n}_{\mathrm{(g)}}~.
   \label{eq: chemisputtering_carbon}
\end{equation}
This kind of chemisputtering has been extensively studied, including studies by~\citet{barlowGraphiteGrainSurface1977} and~\citet{draineChemisputteringInterstellarGraphite1979}. The global process is summarized in the work of~\citet{lenzuniDustEvaporationProtostellar1995}. Since these reactions are dominated by the vaporization process, we use Eq.~\eqref{eq: evaporation_rate_lenzuni} to compute the global radius evolution through hydrogen sputtering, and we use Eq. (24) of~\citet{lenzuniDustEvaporationProtostellar1995} for the yield coefficient:
\begin{equation}
   \indrm{Y}{H} = \indrm{Y}{\max} \exp\left(-\left[\frac{\parallel T - \indrm{T}{\max} \parallel}{\sigma}\right]^{1.75}\right).
\end{equation}
Here $\indrm{Y}{\max} = 3.15\times 10^{-2}$, $\indrm{T}{\max} = 625~\mathrm{K}$, and $\sigma = 1.35~\mathrm{K}$. 

The $\mathrm{H}_2\mathrm{O}$ molecules can also be a source of chemisputtering, as shown by~\citet{stahlerEvolutionProtostarsIII1981}. We again use the same yield coefficient as in~\citet{lenzuniDustEvaporationProtostellar1995} with their Eq.~(27), which is
\begin{equation}
   \indrm{Y}{H_2O} = \indrm{Y}{0} \exp\left(-\frac{T_0}{T}\right),
\end{equation}
with $Y_0 = 6.85\times 10^{5}$ and $T_0 = 29000~\mathrm{K}$.

Finally, the chemisputtering by molecular hydrogen $\mathrm{H}_2$ is also possible. This chemisputtering is not taken into account by~\citet{lenzuniDustEvaporationProtostellar1995} because of the lack of data. In our case we use the  survey of~\citet{krakowskiSURVEYLITERATURECARBONHYDROGEN1970}, which compiles various measurements of the yield coefficient $\indrm{Y}{H_2}$ as a function of the temperature. The results are shown in Fig.~\ref{fig: yield_hydrogen}.

\begin{figure*}[t]
   \centering
   \includegraphics[height = 6.5cm]{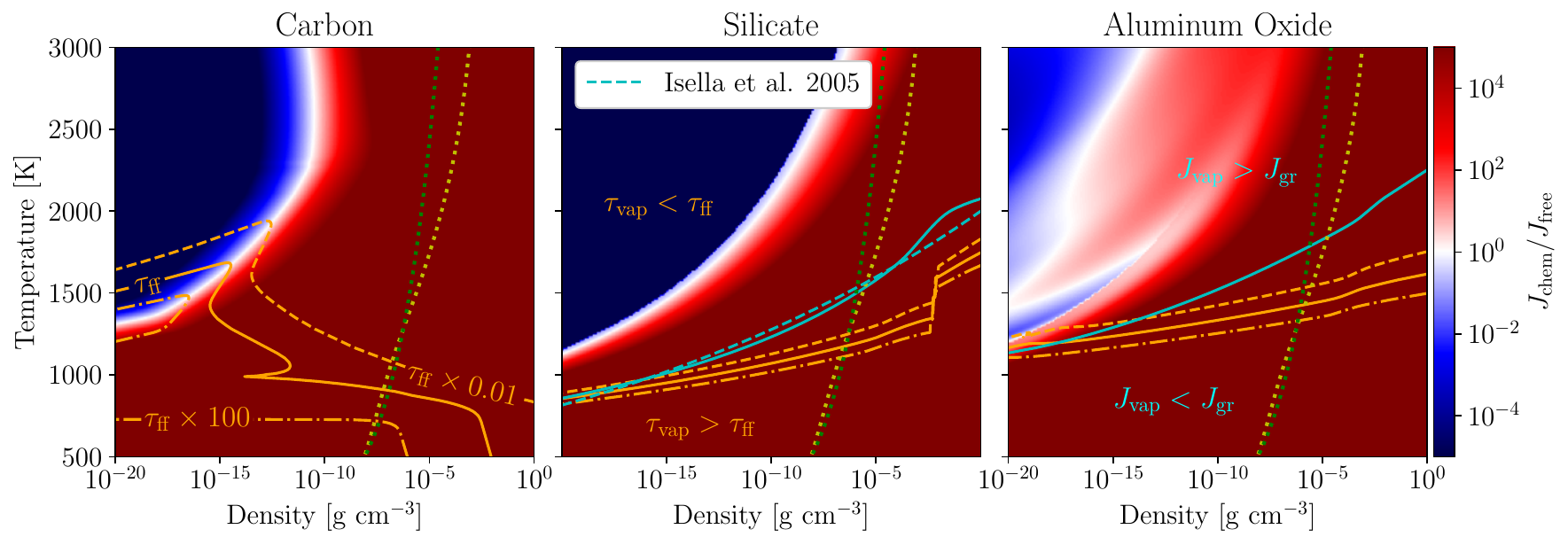}
\caption{Color map showing the ratio between the chemisputtering vaporization and the free vaporization flux as a function of temperature and density of the gas, with a saturation at $10^{-5}$ and $10^{5}$. 
{Red} corresponds to the area where the chemisputtering vaporization flux dominates, and   {blue} corresponds to the area where the free vaporization flux dominates. 
The {yellow dotted line} represents the barotropic law (see Eq.~\eqref{eq: barotropic_law}), and the {green dotted line} represents a typical disk model from~\citet[see our Sect.~\ref{subsec: disk_application} and Eqs.~\eqref{eq: temperature_disk} and~\eqref{eq: density_disk}]{andrewsProtoplanetaryDiskStructures2009}. 
The {cyan solid line} represents the limit where the vaporization flux is equal to the growth flux. The vaporization dominates for higher temperatures (above the line) and the growth dominates for lower temperatures (below the line). The carbon grains do not have this line as vaporization always dominates for the range of temperatures and density of this plot. 
The {cyan dashed line} represents the sublimation limit prescription of the silicates used in~\citet{isellaShapeInnerRim2005}. Its expression is given in Eq.~\eqref{eq: Isela_sublimation_temperature}. 
Finally, the {orange solid line} represents the limit where the vaporization timescale is equal to the free-fall timescale. For higher temperatures (above the line), the vaporization timescale is smaller than the free-fall timescale, and for lower temperatures (below the line), the vaporization timescale is larger than the free-fall timescale. 
The two other {orange lines} represent the limit where the vaporization timescale is equal to the free-fall timescale, but multiplied by a factor of 100 ({dot-dashed}) and 0.01 ({dashed}).
The jump in the {orange lines} on the middle panel is due to a change in the key reactant in reaction~\eqref{eq: chemisputtering_forsterite}, being $\mathrm{Mg}$ for lower densities and $\mathrm{SiO}$ for higher densities.
The vaporization timescales are computed with a grain radius of $5\times 10^{-7}~\mathrm{cm}$ in Eq.~\eqref{eq: characteristic_timescale}.}
\label{fig: timescales_colormap}
\end{figure*}

\begin{figure}[t]
   \centering
   \includegraphics[height = 6.cm]{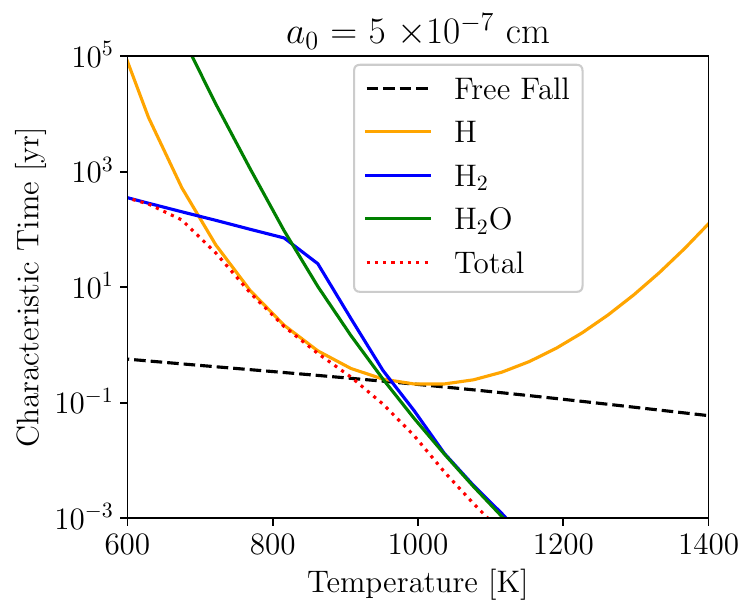}
   \caption{Evolution of the characteristic time of destruction of a carbon grain with a radius of $5\times 10^{-7}~\mathrm{cm}$ as a function of temperature for each reaction. The free-fall time is also shown, assuming a barotropic law for temperature-density relation~\citep{machidaSecondCoreFormation2006}.}
   \label{fig: characteristic_time_carbon}
\end{figure}

%%%%%%%%%%%%%%%%%%%%%%%%%%%%%%%%%%%%%%%%%%%%%%
\subsection{Silicates}
\label{subsec: silicates}

The monomer composing the silicate grains is the group $\mathrm{Mg_2SiO_4}$ (forsterite). 
Several studies have reported experimental data for the vaporization of forsterite in an $\mathrm{H_2}$ atmosphere~\citep{hashimotoEvaporationKineticsForsterite1990,nagaharaVaporizationRateForsterite1994,nagaharaEvaporationForsterite21996,tsuchiyamaEvaporationRatesForsterite1998, kurodaReactionForsteriteHydrogenits2002}. \citet{gailMineralFormationStellar1999} highlighted the study of~\citet{nagaharaEvaporationForsterite21996} and discussed it as a chemical sputtering reaction whereby $\mathrm{H_2}$ reacts with forsterite and enhances the vaporization compared with a vacuum vaporization where there is no reactive gas. 
This dichotomy in reaction mechanisms led these authors to describe the vaporization of olivine with two mechanisms that are summarized by two reactions: the free vaporization
\begin{equation}
   \mathrm{Mg_2SiO_{4}}_{\mathrm{(s)}} \rightleftharpoons 2\mathrm{Mg} + \mathrm{SiO} + \frac{1}{2}\mathrm{O_2}~,
   \label{eq: free_evaporation_forsterite}
\end{equation}
and the chemisputtering by $\mathrm{H_2}$
\begin{equation}
   \mathrm{Mg_2SiO_{4}}_{\mathrm{(s)}} + 3\mathrm{H_2} \rightleftharpoons 2\mathrm{Mg} + \mathrm{SiO} + 3\mathrm{H_2O}~.
   \label{eq: chemisputtering_forsterite}
\end{equation}
For both reactions we use Eq.~\eqref{eq: total_rate} to compute the vaporization and growth rate. The sticking coefficient $\alpha$ is in practice close to $0.1$ for $\mathrm{SiO}$, as shown in~\citet{shornikovMassSpectrometricStudy1998, fedkinVaporPressuresEvaporation2006}, where values between 0.05 and 0.15 have been found.

%%%%%%%%%%%%%%%%%%%%%%%%%%%%%%%%%%%%%%%%%%%%%%
\subsection{Aluminum oxides}
\label{subsec: aluminium_oxide}

The monomer composing the aluminum oxide grains is the group $\mathrm{Al_2O_3}$. In the presence of $\mathrm{H_2}$, there are three main reactions that can occur, shown to be predominant by thermodynamic calculations with the thermochemical software and the database FactSage~\citep{baleFactSageThermochemicalSoftware2016}:
\begin{equation}
   \mathrm{Al_2O_3}_{\mathrm{(s)}} + 3 \mathrm{H_2} \rightleftharpoons 2\mathrm{Al} + 3\mathrm{H_2O}~,
   \label{eq: chemisputtering_aluminium_oxide_1}
\end{equation} 
\begin{equation}
   \mathrm{Al_2O_3}_{\mathrm{(s)}} + 2 \mathrm{H_2} \rightleftharpoons \mathrm{Al_2O} + 2\mathrm{H_2O}~,
   \label{eq: chemisputtering_aluminium_oxide_2}
\end{equation}
\begin{equation}
   \mathrm{Al_2O_3}_{\mathrm{(s)}} + \mathrm{H_2} \rightleftharpoons 2\mathrm{AlO} + 2\mathrm{H_2O}~.
   \label{eq: chemisputtering_aluminium_oxide_3}
\end{equation}
They  are all chemisputtering reactions by $\mathrm{H_2}$. Similarly to the case of silicate, we use Eq.~\eqref{eq: total_rate}.
% with the key species being $\mathrm{Al}$ for reaction~\eqref{eq: chemisputtering_aluminium_oxide_1}, $\mathrm{Al_2O}$ for reaction~\eqref{eq: chemisputtering_aluminium_oxide_2}, and $\mathrm{AlO}$ for reaction~\eqref{eq: chemisputtering_aluminium_oxide_3}.
We can also consider the free vaporization of aluminum oxide, which is described by the reaction
\begin{equation}
   \mathrm{Al_2O_3}_{\mathrm{(s)}} \rightleftharpoons 2\mathrm{Al} + \frac{3}{2}\mathrm{O_2}~.
   \label{eq: free_evaporation_aluminium_oxide}
\end{equation}
The values for the sticking coefficients range between 0.2 and 0.3, according to~\citet{burnsSystematicsEvaporationCoefficient1966}.

%%%%%%%%%%%%%%%%%%%%%%%%%%%%%%%%%%%%%%%%%%%%%%
\subsection{Timescales}
\label{subsec:characteristic_times}

It is possible to compute the characteristic timescale of variation of the grain radius for each of these reactions as a function of temperature. Via Eq.~\eqref{eq: grain_radius_evolution}, we can compute the characteristic time of destruction of a grain of radius $a_0$ by a given reaction as
\begin{equation}
   \indrm{\tau}{grain} = \frac{a_0}{\indrm{V}{m} J_{\mathrm{tot}}}.
   \label{eq: characteristic_timescale}
\end{equation}

This allows us to have a first idea of the importance of each reaction in the evolution of the grain. In the context of  gravitational collapse, the free-fall time is the characteristic time of evolution of the temperature and density of the gas, and is given by
\begin{equation}
   \tau_{\mathrm{ff}} = \sqrt{\frac{3\pi}{32G\rho}}.
\end{equation}
In the context of  gravitational collapse, the barotropic law from~\citet{machidaSecondCoreFormation2006} gives a good idea of the evolution of the temperature as a function of the density. It is defined as
\begin{equation}
   T = T_0\sqrt{1+\left(\frac{\rho}{\rho_1}\right)^{2g_1}}\left(1+\frac{\rho}{\rho_2}\right)^{g_2}\left(1+\frac{\rho}{\rho_3}\right)^{g_3},
   \label{eq: barotropic_law}
\end{equation}
with the critical densities $\rho_1 = 3.9\times 10^{-13}~\mathrm{g~cm^{-3}}$, $\rho_2 = 3.9\times 10^{-8}~\mathrm{g~cm^{-3}}$, and $\rho_3 = 3.9\times 10^{-3}~\mathrm{g~cm^{-3}}$, and the coefficients $g_1 = 0.4$, $g_2 = -0.3$, and $g_3 = 0.56667$.

From this we can obtain a good idea of the sublimation limit of each grain material by examining Fig.~\ref{fig: timescales_colormap}. We note that, in the context of  gravitational collapse (dotted black line), chemisputtering dominates vaporization for all three types of grain. 
Then, there are two conditions that determine if a dust grain is sublimated for a given temperature and density: the vaporization flux must be greater than the growth flux, such that vaporization dominates growth (the vaporization-growth limit), and the vaporization timescale must be smaller than the dynamical timescale of the gas phase (the dynamical limit), so that the grain has time to evaporate at this temperature. In the context of increasing temperature, the sublimation limit for each material in Fig.~\ref{fig: timescales_colormap} corresponds to the higher curve between the orange and black solid lines, representing the dynamical limit and the vaporization-growth limit, respectively.

These considerations demonstrate that in the context of gravitational collapse (black dotted line), the vaporization of silicate and aluminum oxide grains is determined by the vaporization-growth limit. We see that our model reproduces quite well the silicate sublimation limit used in \citet[the blue dashed line]{isellaShapeInnerRim2005}, which is based on observational data from~\citet{pollackCompositionRadiativeProperties1994} and gives the sublimation temperature of silicate grains to be equal to
\begin{equation}
   T_{\mathrm{sub}} = 2000\left(\frac{\indrm{\rho}{gas}}{1~\mathrm{g~cm^{-3}}}\right)^{1.95\times 10^{-2}}~\mathrm{K}.
   \label{eq: Isela_sublimation_temperature}
\end{equation}
For carbon grains, we see that the dynamical limit is the relevant factor for the sublimation limit. We also observe that if we change the value of the dynamical timescale by a factor of 0.01 and 100, it significantly changes the sublimation limit of the carbon grains, highlighting the crucial role of the dynamical evolution of the gas in determining the evolution of carbon grains, while it only slightly affects the dynamical limit of silicate and aluminum oxide grains.

To understand why the carbon material is very dependent on the dynamical evolution of the gas, we can examine the effects of each reaction that can destroy a carbon grain.
In Fig.~\ref{fig: characteristic_time_carbon} we compare the free-fall time and different characteristic times of destruction of a carbon grain with a radius of $5\times 10^{-7}~\mathrm{cm}$ (the smallest grains in our size distribution) as a function of temperature, assuming a barotropic law.
We observe that the chemisputtering by $\mathrm{H}$ atoms has a range of temperatures where the characteristic time has approximately the same order of magnitude as the free-fall time, between 900~K and 1100~K. This has the consequence of reducing the steepness of the total characteristic vaporization time, resulting in significant changes in the crossing temperature with the different characteristic timescales. This suggests that the dynamical approach is necessary to accurately model the evolution of carbon grains.

%%%%%%%%%%%%%%%%%%%%%%%%%%%%%%%%%%%%%%%%%%%%%%%%%%%%%%%%%%%%%%%%%%%%%%%%%%%%%%%%%%%%%%%%%%%%
%%%%%%%%%%%%%%%%%%%%%%%%%%%%%%%%%%%%%%%%%%%%%%%%%%%%%%%%%%%%%%%%%%%%%%%%%%%%%%%%%%%%%%%%%%%%
\section{Dust evolution computation}
\label{sec: dust_evolution_computation}

\begin{figure}[t]
   \centering
   \includegraphics[height = 6cm]{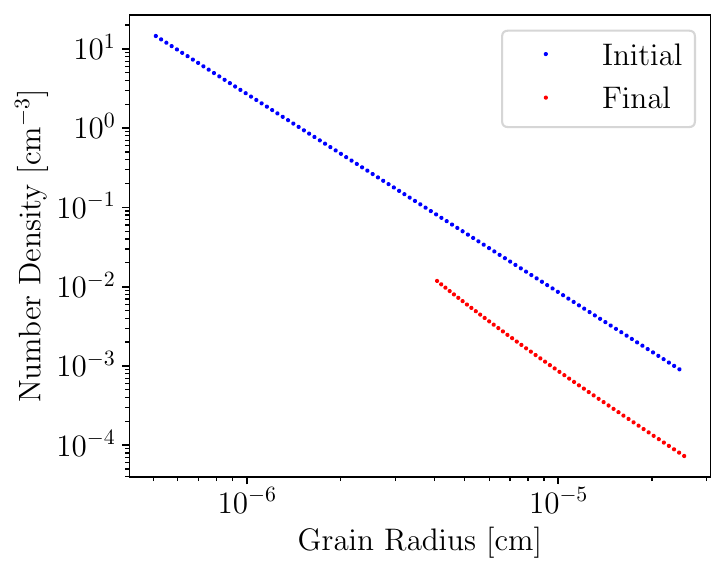}
   \caption{Dust size distribution for aluminum oxide grains at the beginning of the computation (in {blue}) and at the end of the computation (in {red}) for the trajectory from~\citet{bhandareMixingEasyNew2024a} shown in Fig.~\ref{fig: Asmita_trajectories}.}
   \label{fig: Asmita_size_distribution}
\end{figure}

Now that we have defined the content of our dust grains, we can try to apply our model to different temperature and density evolutions $T(t)$ and $\rho(t)$, which we call trajectories from now on.

%%%%%%%%%%%%%%%%%%%%%%%%%%%%%%%%%%%%%%%%%%%%%%
\subsection{Initial setup}
\label{subsec: size_distribution}

Before computing the evolution of the dust grains, we need to set the initial conditions for the quantity of  dust grains  and the size distribution. For the quantity of each material, we use the values of the dust-to-gas ratios in Table~\ref{tab: dust_materials} to compute their total density from the initial gas density of the trajectory.

For the size distribution, we use the power-law distribution from \citet{mathisSizeDistributionInterstellar1977}, also known as the MRN distribution. According to the MRN distribution, between $a_{\min} = 5 \times 10^{-7}~\mathrm{cm}$ and $a_{\max} = 2.5 \times 10^{-5}~\mathrm{cm}$, the grain size distribution is  nonzero and can be expressed as
\begin{equation}
   \mathrm{d}n(a) = C a^{\lambda} \mathrm{d}a,
   \label{eq: mrn_distribution}
\end{equation}
where $\mathrm{d}n$ is the number density of grains that have a radius contained between $a$ and $a+\mathrm{d}a$, $\lambda$ is the power-law index and is close to $-3.5$ according to \citet{mathisSizeDistributionInterstellar1977}, and $C$ is a normalization constant. This constant can be computed from the total dust density $\rho_{\mathrm{dust}}$ as
\begin{equation}
   \rho_{\mathrm{dust}} = \int_{a_{\min}}^{a_{\max}} \indrm{\rho}{grain} \frac{4}{3}\pi a^3 C a^{\lambda} \mathrm{d}a.
\end{equation}
The computation of this integral gives the expression of $C$,
\begin{equation}
   C = \frac{\rho_{\mathrm{dust}}}{\rho_{\mathrm{grain}}}\frac{3(\lambda+4)}{4\pi a_{\min}^{\lambda+4}(\xi^{\lambda+4} - 1)},
\end{equation}
where $\xi = a_{\max}/a_{\min}$. To compute the evolution of the grain size distribution, and also the  non-ideal MHD resistivities and the opacities, it is useful to work directly with a discretized version of the MRN distribution. To do this, we choose a number of bins $\indrm{N}{bin}$ and we define the radius limit between the bin $i$ and $i+1$ (where $i$ is between 1 and $\indrm{N}{bin}$) as $a_{\mathrm{lim},i} = a_{\min} \xi^{i/\indrm{N}{bin}}$, so that they are logarithmically spaced. The number of grains in bin $i$ is then given by
\begin{equation}
   n_i = \frac{C}{\lambda+1}\left(a_{\mathrm{lim},i}^{\lambda+1} - a_{\mathrm{lim},i-1}^{\lambda+1}\right).
\end{equation}
We now choose that all grains in each bin have the same radius $a_i$. The choice of this radius $a_i$ depends on whether we want to conserve the total mass of the grains or their total surface area. However, if we take a sufficiently high number of bins, the difference between the two choices becomes negligible, and we can simply choose the geometric mean of the bin, so that $a_i = \sqrt{a_{\mathrm{lim},i}a_{\mathrm{lim},i-1}}$. In our case, we set $\indrm{N}{bin} = 100$, such that the error in the total dust mass and the total surface area of the MRN distribution is less than $0.04\%$. This process results in the grain repartition represented by the blue dot in Fig.~\ref{fig: Asmita_size_distribution}. We can now apply the method described in Sect.~\ref{subsec: size_distribution} for the evolution of each radius in the distribution.

%%%%%%%%%%%%%%%%%%%%%%%%%%%%%%%%%%%%%%%%%%%%%%
\subsection{Dust evolution}
\label{subsec: dust_evolution}

\begin{figure*}[t]
   \begin{subfigure}{0.28\textwidth}
      \centering
      \includegraphics[height = 6.cm]{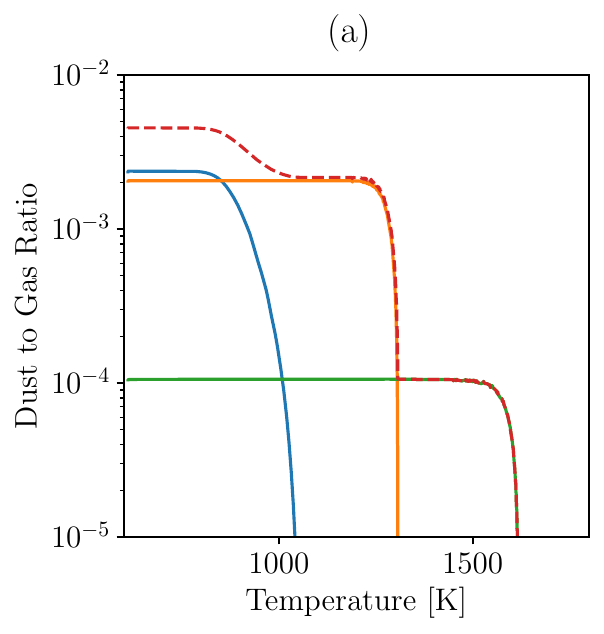}
   \end{subfigure}
   \hfill 
\begin{subfigure}{0.28\textwidth}
   \centering
   \includegraphics[height = 6.cm]{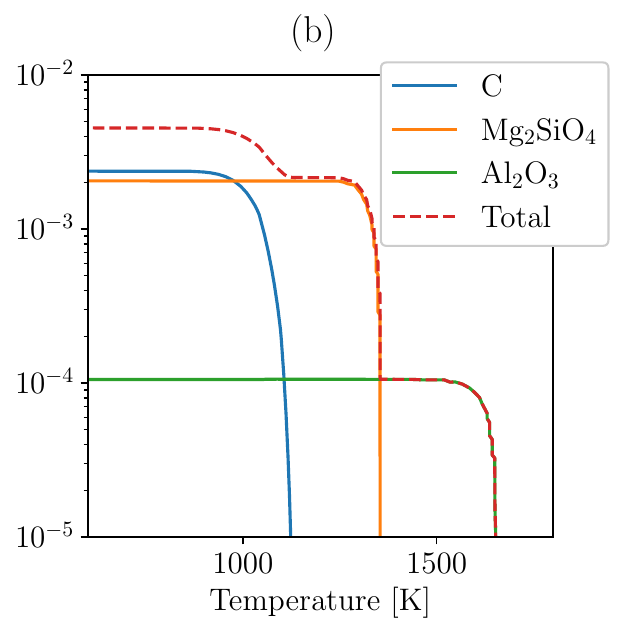}

\end{subfigure}
\hfill 
\begin{subfigure}{0.31\textwidth}
   \centering
   \includegraphics[height = 6.cm]{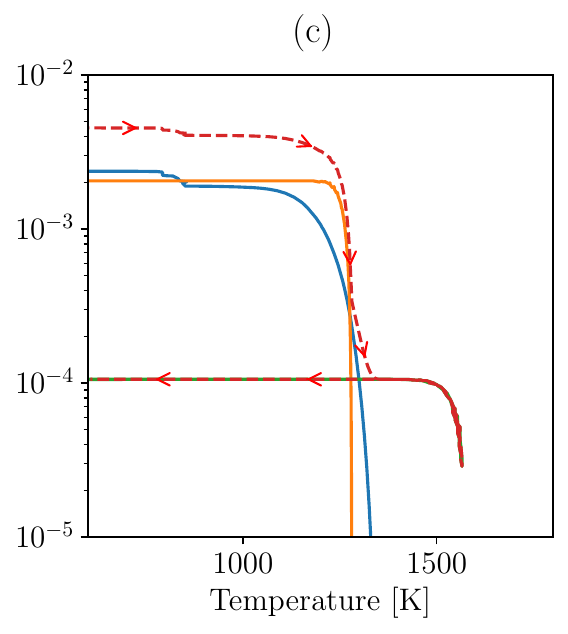}

\end{subfigure}
\caption{Dust-to-gas ratio as a function of temperature for three different trajectories. For panel (a) we use the trajectory from \citet{lenzuniDustEvaporationProtostellar1995}, shown in Fig.~\ref{fig: lenzuni_vs_ramses}. For panel (b) we use a trajectory from a collapse simulation, also shown in Fig.~\ref{fig: lenzuni_vs_ramses}. For panel (c) we use a trajectory from the collapse simulation of \citet{bhandareMixingEasyNew2024a}, shown in Fig.~\ref{fig: Asmita_trajectories}. The arrows in the right panel indicate the time evolution of the system.}
\label{fig: dust_evolution}
\end{figure*}

\begin{figure}[]
   \centering
   \includegraphics[height = 12.cm]{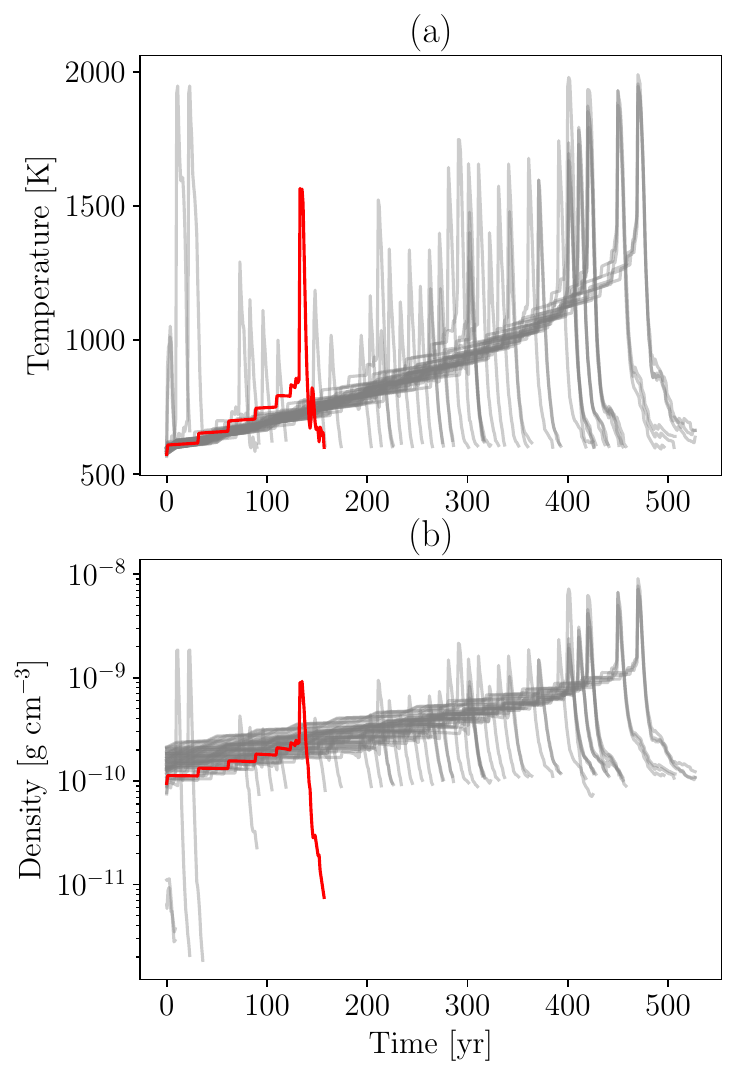}
   \caption{Evolutions of the temperature in panel (a) and the density in panel (b) of the gas surrounding one-micron grains in the collapse simulation of \citet{bhandareMixingEasyNew2024a}. The 55 trajectories selected from the simulation are those where the grains experience a maximum temperature between 1000~K and 2000~K. Each trajectory starts at a temperature of $600~\mathrm{K}$. The {red solid line} represents the trajectory chosen in this article to compute the dust evolution. The grain position evolution is shown in Fig.~\ref{fig: Asmita_trajectories_position}.}
   \label{fig: Asmita_trajectories}
\end{figure}
\begin{figure}[]
   \centering
   \includegraphics[height = 6.cm]{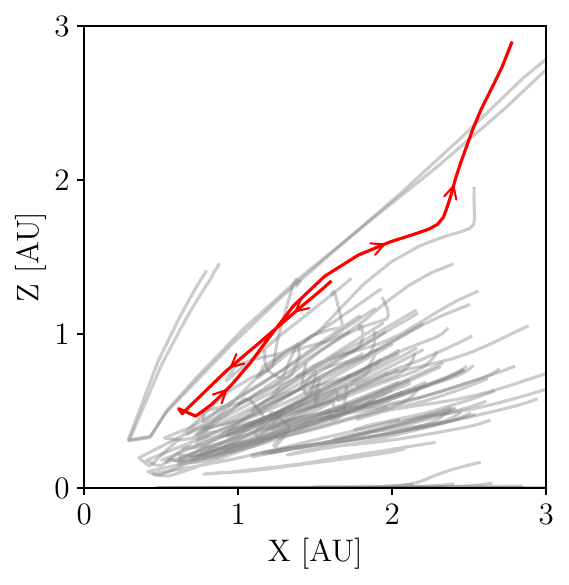}
   \caption{Evolution of the grain positions associated with the temperature and density evolution from panel (a) and (b) of Fig.~\ref{fig: Asmita_trajectories}.}
   \label{fig: Asmita_trajectories_position}
\end{figure}

\begin{figure*}[]
   \centering
   \includegraphics[height = 6.cm]{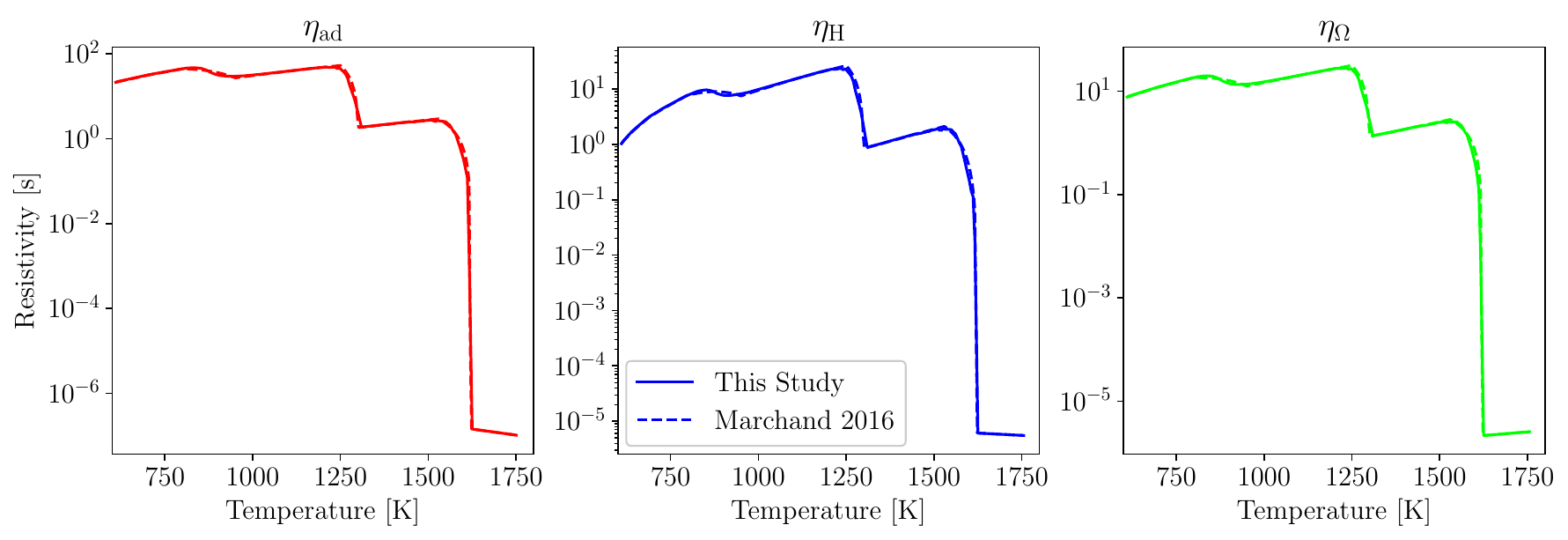}
   \caption{Evolutions of the three non-ideal MHD resistivities as a function of temperature for the trajectory from \citet{lenzuniDustEvaporationProtostellar1995} shown in Fig.~\ref{fig: lenzuni_vs_ramses}. On the left is shown the ambipolar diffusion resistivity $\indrm{\eta}{ad}$, in the middle is the Hall effect resistivity $\indrm{\eta}{H}$, and on the right is the Ohmic resistivity $\indrm{\eta}{\Omega}$. In each plot the resistivity computed through the dust evolution computation is shown as a {solid line}, and the resistivity computed through the~\citet{marchandChemicalSolverCompute2016} vaporization modeling is shown as a {dashed line}.}
   \label{fig: resistivity_lenzuni_evolution_comparison}
\end{figure*}

\begin{figure*}[]
   \centering
   \includegraphics[height = 6.cm]{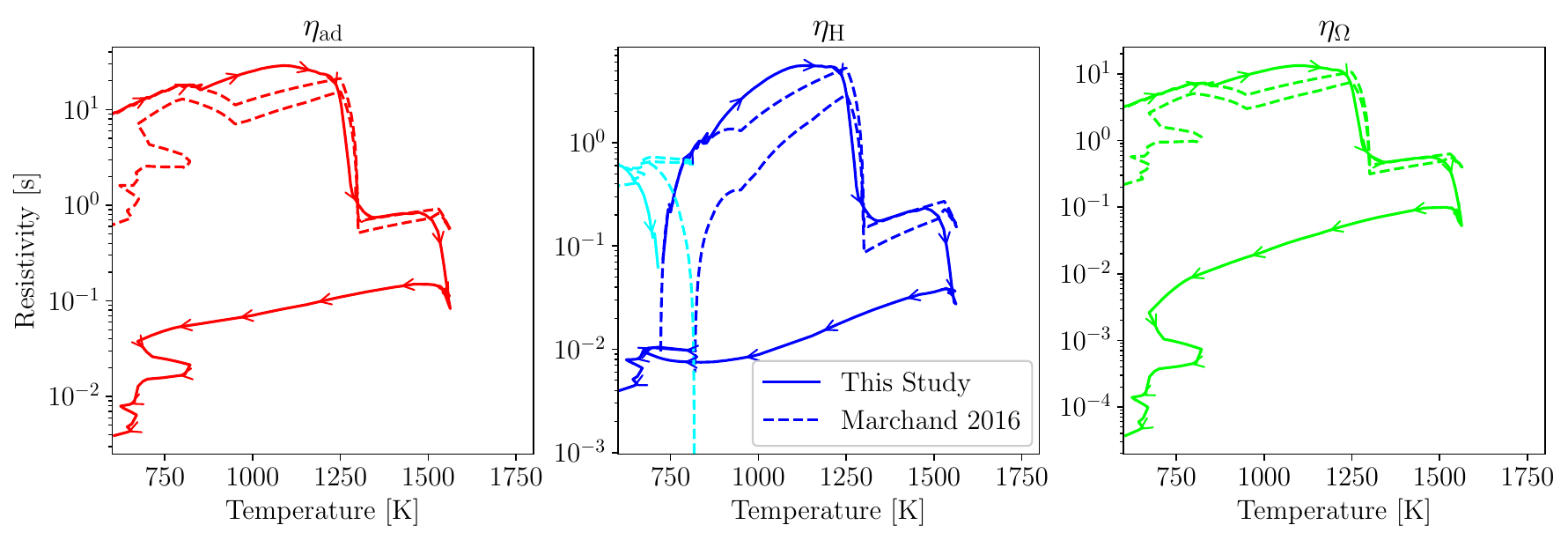}
   \caption{Same as Fig.~\ref{fig: resistivity_lenzuni_evolution_comparison}, but for the trajectory from~\citet{bhandareMixingEasyNew2024a} shown in Fig.~\ref{fig: Asmita_trajectories}. The arrows indicate the time evolution of the system.}
   \label{fig: resistivity_asmita_evolution_comparison}
\end{figure*}

\begin{figure*}[t]
   \begin{subfigure}{0.29\textwidth}
      \centering
      \includegraphics[height = 6.cm]{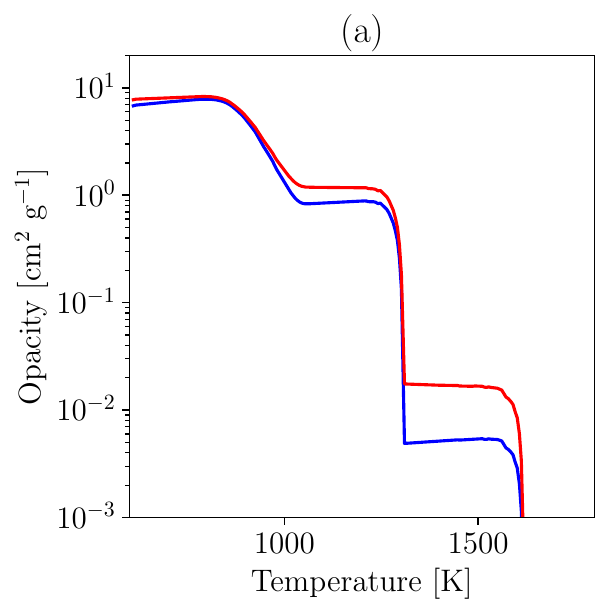}
   \end{subfigure}
   \hfill 
\begin{subfigure}{0.29\textwidth}
   \centering
   \includegraphics[height = 6.cm]{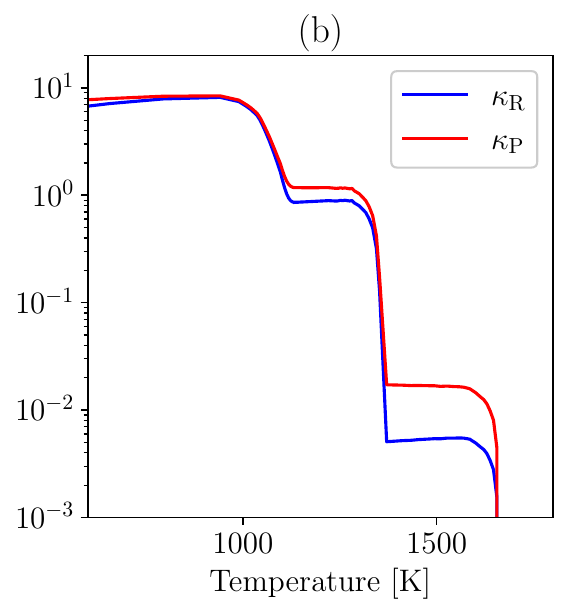}

\end{subfigure}
\hfill 
\begin{subfigure}{0.29\textwidth}
   \centering
   \includegraphics[height = 6.cm]{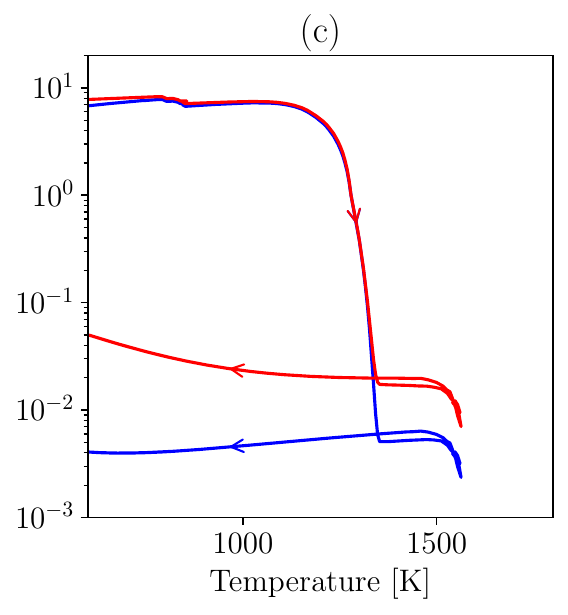}

\end{subfigure}
\caption{Planck and Rosseland mean opacities as a function of temperature for three different trajectories. For panel (a) we use the trajectory from \citet{lenzuniDustEvaporationProtostellar1995}, shown in Fig.~\ref{fig: lenzuni_vs_ramses}. For panel (b) we use a trajectory from a collapse simulation, also shown in Fig.~\ref{fig: lenzuni_vs_ramses}. For panel (c) we use a trajectory from the collapse simulation of~\citet{bhandareMixingEasyNew2024a} shown in  Fig.~\ref{fig: Asmita_trajectories}.}
\label{fig: opacities_evolution}
\end{figure*}

We can now compute the evolution of the MRN dust distribution for a given trajectory, which consists of a temperature $T(t)$ and density $\rho(t)$ evolution. We show the results of the computation for three different trajectories. The first one is the trajectory from \citet{lenzuniDustEvaporationProtostellar1995}, which is shown in Fig.~\ref{fig: lenzuni_vs_ramses}. 

The second one is a trajectory that follows the evolution of the central computational cell of a collapse simulation performed with the AMR hydrodynamic code RAMSES~\citep{teyssierCosmologicalHydrodynamicsAdaptive2002}. The numerical and physical setup is similar to the one used in~\citet{ahmadBirthEarlyEvolution2023}: we start from a one solar mass cloud with a ratio of thermal to gravitational energy of $\alpha = 0.25$, without any rotation or magnetic field. The evolution in temperature and density of the central cell is shown in Fig.~\ref{fig: lenzuni_vs_ramses}.

The third is the trajectory of a one-micron grain extracted from a simulation of the collapse of a solar mass  molecular cloud performed by~\citet{bhandareMixingEasyNew2024a}, shown in Fig.~\ref{fig: Asmita_trajectories}. The grain trajectories selected here are the ones that experience an ejection from the center of the system (see Fig.~\ref{fig: Asmita_trajectories_position}), causing a fast decrease in temperature and density. The evolution of the dust-to-gas ratio for each grain species and for each trajectory is shown in Fig.~\ref{fig: dust_evolution}.

First of all, we see that the results for the \citet{lenzuniDustEvaporationProtostellar1995} trajectory (left panel in Fig.~\ref{fig: dust_evolution}) are in good agreement with what was obtained in their article. The only difference is the temperature of full destruction of aluminum oxide, which is around 1610~K in our computation, whereas it is around 1720~K for~\citet{lenzuniDustEvaporationProtostellar1995}. This difference probably arises from the difference in the modeling of the vaporization process. 
In our case we use a set of three reactions~\eqref{eq: chemisputtering_aluminium_oxide_1},~\eqref{eq: chemisputtering_aluminium_oxide_2}, and~\eqref{eq: chemisputtering_aluminium_oxide_3}, while in their case they computed the chemical equilibrium of $\mathrm{Al_2O_3}_{\mathrm{(s)}}$ between the chemical species $\mathrm{Al}$, $\mathrm{AlOH}$, $\mathrm{Al_2O}$, $\mathrm{AlH}$, $\mathrm{H_2}$, and $\mathrm{H_2O}$.

The main difference between Lenzuni's trajectory and the trajectory from the collapse simulation is the speed of evolution of temperature and density (see Fig.~\ref{fig: lenzuni_vs_ramses}), where the latter evolves much faster.
This difference directly impacts the results of the computation, particularly for the carbonaceous grains. Since the vaporization reactions for this type of grain have a timescale close to the evolution of temperature, as shown in Fig.~\ref{fig: characteristic_time_carbon}, a change in the speed of temperature variation modifies the temperature of full destruction of the carbonaceous grains. For Lenzuni's trajectory it is around 1030~K, while it becomes around 1120~K for the collapse simulation.
There is no real impact on the other two species of grains since the timescales for their reactions are much shorter than the evolution of temperature.

In the case of the trajectory from the simulation of \citet{bhandareMixingEasyNew2024a} shown in Fig.~\ref{fig: Asmita_trajectories}, there are two regimes. It starts with the temperature and density increasing at a rate similar to the collapse simulation, but at a certain point, the growth rate suddenly increases until it reaches a maximum, and then it decreases at the same rate (caused by the ejection of the grains from the center of the collapse).
The extremely fast increase in temperature implies a shift of the temperature of full destruction of the carbonaceous grains to around 1310~K (higher than the vaporization limit of silicates, which is around 1290~K). The subsequent decrease in temperature before destroying all the aluminum oxide grains leads to a sudden growth of the latter grains, recovering all their mass when the temperature is sufficiently low. However, as the carbonaceous grains and silicates have already been completely evaporated, their population remains at zero, resulting in a drastic change in the dust quantity for temperatures lower than 1300~K compared to before the temperature spike. 
This result is due to the absence of nucleation in our model, and the impossibility of recondensation of carbonaceous and silicate materials on the remaining aluminum oxide grains, which is discussed in Sect.~\ref{subsec: limitations}.
The recondensation of the aluminum oxide grains is also visible in the size distribution shown in Fig.~\ref{fig: Asmita_size_distribution}. As the recondensation can only happen on the grains that are still present in the gas, we obtain at the end of the trajectories a size distribution that is devoid of its smallest grains.

For all the other trajectories shown in Fig.~\ref{fig: Asmita_trajectories}, the results are similar and mainly depend on the maximum temperature reached. If this temperature is below 1300~K, all types of grains recover their initial mass. If it is between 1300~K and 1600~K, the results are similar to what was just discussed. However, if the temperature exceeds 1600~K, the grains are completely destroyed and there is no recondensation when the temperature decreases.

%%%%%%%%%%%%%%%%%%%%%%%%%%%%%%%%%%%%%%%%%%%%%%
\subsection{Non-ideal MHD resistivities}
\label{subsec: non_ideal_mhd_resistivities}

The three non-ideal MHD resistivities that appear in the non-ideal MHD equations~\citep{marchandChemicalSolverCompute2016} can be computed from a given discretized grain size distribution. For this computation, we use a routine using the analytical derivation of the grain ionization
from~\citet{marchandFastMethodsTracking2021}. This routine needs as input a discretized grain size distribution, the temperature $T$, the numerical density of particles in the gas phase $\indrm{n}{gas}$, the ionization rate $\indrm{x}{i}$, and the magnetic field magnitude $B$. The temperature is given directly by the trajectory we used to compute the dust vaporization, and the numerical density of particles in the gas phase is given by the gas density and the mean molecular weight of the gas $\indrm{\mu}{gas}$, so that $\indrm{n}{gas} = \rho/\left(\indrm{\mu}{gas}\indrm{m}{H}\right)$, with $\indrm{\mu}{gas} = 2.34$ for a solar composition, and $\indrm{m}{H}$ is the mass of hydrogen. For the ionization rate and the magnetic field magnitude we use the same prescription as~\citet{marchandChemicalSolverCompute2016}, we set a constant ionization rate $\indrm{x}{i} = 10^{-17}~\mathrm{s^{-1}}$, and we follow~\citet{liNonidealMHDEffects2011} such that the magnetic field scales as
\begin{equation}
   B = 1.43\times 10^{-7}\sqrt{\frac{\indrm{n}{gas}}{1~\mathrm{cm^{-3}}}}~\mathrm{G}. 
\end{equation}

% $a_{\min} = 1.81\times 10^{-6}~\mathrm{cm}$ and $a_{\max} = 9.049\times 10^{-5}~\mathrm{cm}$, following~\citet{kunzNonisothermalStageMagnetic2009}.

The goal of this section is to compare the results of our vaporization computation with those of the vaporization modeling by~\citet{marchandChemicalSolverCompute2016}. In this article it is assumed that the grain sizes always follow an MRN distribution, and it is assumed that the grain quantity is a function only of the temperature: each type of grain evaporates linearly within a given temperature range. These ranges were chosen based on the results of~\citet{lenzuniDustEvaporationProtostellar1995}:  carbon evaporates between 750~K and 1100~K, silicates between 1200~K and 1300~K, and aluminum oxides between 1600~K and 1700~K. As our computation gives slightly different results for   Lenzuni's trajectory, we chose to modify the vaporization temperature ranges so that the strict comparison of the two resistivity models is meaningful. We set the vaporization temperature ranges to be between 800~K and 950~K for the carbon grains, between 1250~K and 1300~K for the silicate grains, and between 1530~K and 1620~K for the aluminum oxide grains.
With this choice, we observe in Fig.~\ref{fig: resistivity_lenzuni_evolution_comparison} that the resistivities computed from our vaporization computation for Lenzuni's trajectory are almost identical to the resistivities computed with the simple vaporization model described above. 

However, if we maintain the same vaporization model and consider the \citet{bhandareMixingEasyNew2024a} simulation trajectory, we obersve some differences in our computations (see Fig.~\ref{fig: resistivity_asmita_evolution_comparison}). 
First of all, the late vaporization of the carbonaceous grains generates higher resistivities for temperatures between 950~K and 1300~K. When the temperature decreases below 1300~K, the resistivities are much lower (by a factor of 100) than the simple vaporization model. In the~\citet{marchandChemicalSolverCompute2016} vaporization model, all the grains recover their mass when the temperature decreases sufficiently, which does not occur in our computation since the materials cannot recondense if no more grain is present. Additionally, the resistivities between 1300~K and 1550~K just after the spike in temperatures are also different by a factor of 10 between the model and the computation, while the aluminum oxide grains recovered their initial mass (see Fig.~\ref{fig: dust_evolution}). The reason for this is shown in Fig.~\ref{fig: Asmita_size_distribution}: the spike in temperature removes the smallest grains. Even if the total mass of the aluminum oxide grains is recovered, this causes the grain number to be reduced, thereby decreasing the resistivities.

%%%%%%%%%%%%%%%%%%%%%%%%%%%%%%%%%%%%%%%%%%%%%%
\subsection{Opacities}
\label{subsec: opacities}

It is also possible to compute the opacities of a given dust distribution. For this purpose, we use the library DSHARP~\citep{birnstielDiskSubstructuresHigh2018}. By taking a grain radius $a$ and the complex refractive index $\underbar{n} = n + ik$ of the material forming the grain, it can compute for a given frequency $f$ the opacity per unit of mass $\kappa(f, a)$.
For a mixture of $N_{\mathrm{species}}$ grain species with $N_\mathrm{radii}$ different radii, the total opacity per unit of mass is given by
\begin{equation}
   \indrm{\kappa}{tot}(f) = \sum_{i}^{N_{\mathrm{species}}} \frac{\sum_{j}^{N_\mathrm{radii}} n_{i,j} a_j^3 \kappa_i(f, a_j)}{\sum_{j}^{N_\mathrm{radii}} n_{i,j} a_j^3},
   \label{eq: total_opacity}
\end{equation}
with $n_{i,j}$ the number density of grains of species $i$ and radius $j$.

There are several ways to compute a mean opacity over the frequency. The two most common are the Planck and Rosseland mean opacities, which appear in the equations for the radiative transfer and are then important for the modeling of the protostar formation. The Planck opacity is defined as
\begin{equation}
   \kappa_{\mathrm{P}} = \frac{\int \kappa(f) B(f, T) \mathrm{d}f}{\int B(f, T) \mathrm{d}f},
   \label{eq: planck_opacity}
\end{equation}
where $B\left(f, T\right)$ is the Planck function. The Rosseland opacity is defined as
\begin{equation}
   \frac{1}{\kappa_{\mathrm{R}}} = \frac{\int \frac{1}{\kappa(f)} \frac{\partial B(f, T)}{\partial T} \mathrm{d}f}{\int \frac{\partial B(f, T)}{\partial T} \mathrm{d}f}.
   \label{eq: rosseland_opacity}
\end{equation}

We can track the evolution of the Planck and Rosseland mean opacities for each trajectory. For the values of the complex refractive index, we use the values from \citet{zubkoOpticalConstantsCosmic1996} for carbonaceous grains, \citet{draineScatteringInterstellarDust2003} for silicate grains, and \citet{erikssonInfraredOpticalProperties1981} and \citet{kischkatMidinfraredOpticalProperties2012} for aluminum oxide grains. The results are shown in Fig.~\ref{fig: opacities_evolution}. We observe that each dust vaporization phase reduces the total opacity of the medium. There are not many differences between the~\citet{lenzuniDustEvaporationProtostellar1995} trajectory and the collapse simulation trajectory, except for a shift in the destruction of carbonaceous grains to higher temperatures, which also leads to a slight shift in the opacities. For the~\citet{bhandareMixingEasyNew2024a} simulation trajectory, the simultaneous vaporization of carbonaceous and silicate grains results in a significant drop in the opacities around 1300~K. The recondensation of aluminum oxide grains as the temperature starts to decrease leads to a slight increase in the opacities, as the radius of surviving grains increases.

%%%%%%%%%%%%%%%%%%%%%%%%%%%%%%%%%%%%%%%%%%%%%%%%%%%%%%%%%%%%%%%%%%%%%%%%%%%%%%%%%%%%%%%%%%%%
%%%%%%%%%%%%%%%%%%%%%%%%%%%%%%%%%%%%%%%%%%%%%%%%%%%%%%%%%%%%%%%%%%%%%%%%%%%%%%%%%%%%%%%%%%%%
\section{Discussion}
\label{sec: discussion}

%%%%%%%%%%%%%%%%%%%%%%%%%%%%%%%%%%%%%%%%%%%%%%
\subsection{Limitations}
\label{subsec: limitations}

The main limitations that we considered for this study are as follows:
\begin{enumerate}
   \item We limited the study to pure grain, implying that recondensation of a given material is not possible on other kinds of grains. It could also change the vaporization process since a given material could be protected from chemisputtering or free vaporization by another material covering it, one  that evaporates at higher temperatures.
   \item We did not take into account   nucleation, meaning that we did not create any new grains. This could potentially allow fully destroyed species to start to recondense. 
   \item We supposed that the gas phase is in chemical equilibrium, which is probably not the case. 
   It is notably argued in~\citet{lenzuniDustEvaporationProtostellar1995} that the production of CO molecules from CH$_n$ molecules, product of the chemisputtering of the carbonaceous grains, is too slow to happen during the grain vaporization process.
   This kinetic limitation allows the carbon atoms to recondense more easily on the grains, shifting upward the temperature of full destruction of carbonaceous grains. This could be corrected in the future by computing the chemical evolution of the gas phase, with the code ULCHEM for instance~\citep{holdshipUCLCHEMGasgrainChemical2017}.
   \item The model used to compute non-ideal MHD resistivities~\citep{marchandFastMethodsTracking2021} is different from the model developed in \citet{marchandChemicalSolverCompute2016}, which is used to compute the resistivity table used in numerical simulations~\citep{vaytetProtostellarBirthAmbipolar2018}. 
   The main reasons for these differences are that grain-grain collisions and thermionic emission (significant from ~800 K) are not taken into account in the \citet{marchandFastMethodsTracking2021} model, and the resistivity table from~\citet{marchandChemicalSolverCompute2016} is limited to grains with only one charge for its computation. These three effects imply that, in disk regions, the table gives resistivity values that are lower than in the model used in this paper. However, these differences do not change the global results and conclusions of Sect.~\ref{subsec: non_ideal_mhd_resistivities}.
   \item We do not know how the quantity of dust materials that has been ejected and experienced partial sublimation compares with the remaining dust in the disk. As  the ejected dust goes back into the outer regions of the  disk \citep{tsukamotoAshfallInducedMolecular2021}, this could potentially change  the MHD resistivities if the dust reflux from the ejection  is not negligible compared to the remaining dust in the envelope.
\end{enumerate}

%%%%%%%%%%%%%%%%%%%%%%%%%%%%%%%%%%%%%%%%%%%%%%
\subsection{Disk application}
\label{subsec: disk_application}

\begin{figure}[b]
   \centering
   \includegraphics[height = 6cm]{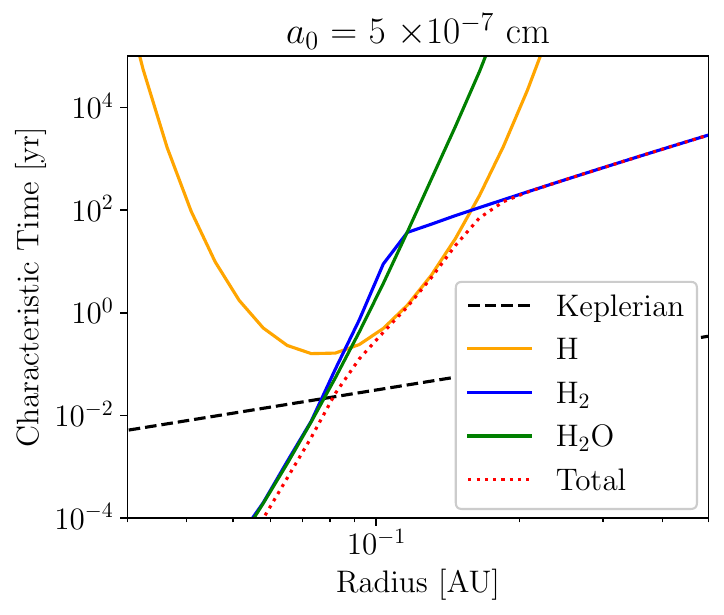}
   \caption{Evolution of the characteristic time of destruction of a carbon grain with a radius of $5\times 10^{-7}~\mathrm{cm}$ as a function of the radius for each reaction, for the~\citet{andrewsProtoplanetaryDiskStructures2009} disk model. The Keplerian time is also shown for comparison.}
   \label{fig: characteristic_time_carbon_disk}
\end{figure}

Another context where we could apply our grain evolutionary model could be the protoplanetary disks. The main differences with the protostellar collapse is the characteristic timescale of evolution of the gas temperature and density: it is not the free-fall time but the Keplerian time, which corresponds to the time to perform a full rotation around the central star. The Keplerian time is given by the Kepler's third law and is expressed as
\begin{equation}
   \tau_{\mathrm{K}} = \sqrt{\frac{4\pi^2 r^3}{G M_{\star}}},
\end{equation}
with $M_{\star}$ the mass of the central star and $r$ the distance to the central star. We can perform the same timescale analysis as in Sect.~\ref{subsec:characteristic_times} for the protostellar disk. We use for that a typical disk model orbiting a one solar mass central star from~\citet{andrewsProtoplanetaryDiskStructures2009} which is in good agreement with observations, {and gives the following expression for temperature and middle density of the disk}
\begin{equation}
   T = 280 \left(\frac{r}{1~\mathrm{AU}}\right)^{-1}~\mathrm{K},
   \label{eq: temperature_disk}
\end{equation}
\begin{equation}
   \rho = 6\times 10^{-10} \left(\frac{r}{1~\mathrm{AU}}\right)^{-9/4}~\mathrm{g~cm^{-3}}.
   \label{eq: density_disk}
\end{equation}
The temperature-density implicit relation is shown in Fig.~\ref{fig: timescales_colormap} (dotted green line). We see that the curve is quite close to the barotropic law for temperature lower than 1600~K, such that the sublimation limit for silicate and aluminum oxide grains should remain the same as in our collapse application. However, the carbonaceous grains are affected by the dynamical  evolution of the gas, and the  Keplerian time is much longer than the free-fall time so we could expect a difference in the sublimation limit.

It is possible again to compute the timescale of destruction of a carbon grain of radius $a_0$ for each kind of reaction. The results for the smallest grain of the MRN distribution are shown in Fig.~\ref{fig: characteristic_time_carbon_disk}, with a comparison with the Keplerian time. The crossing of the total timescale (red dotted line) with the Keplerian time curve gives us an idea of the  radius at which carbon grains should be destroyed. The chemisputtering by $\mathrm{H}_2$ and $\mathrm{H_2O}$ give us a threshold radius of around $8\times 10^{-2}~\mathrm{AU}$.

The chemisputtering by $\mathrm{H_2}$ has a radius range of   around $1\times 10^{-1}~\mathrm{AU}$, where it is the main reaction destroying the grains, and it is a factor of 10 longer than the Keplerian time. This has the consequence of reducing the steepness of the total vaporization timescale. As a result, there is a larger range of radii around $1\times 10^{-1}~\mathrm{AU}$ where the total vaporization timescale is greater than the Keplerian time, but only by a factor of 10 to 100. This means that in a few revolutions, the carbon grains can be greatly affected by the H-sputtering when they are in this range of radii from the central stars. This suggests that a dynamical approach to the vaporization of carbon grains in the protoplanetary disk may be necessary as the H-sputtering prevents us from determining a specific radius of destruction.

%%%%%%%%%%%%%%%%%%%%%%%%%%%%%%%%%%%%%%%%%%%%%%%%%%%%%%%%%%%%%%%%%%%%%%%%%%%%%%%%%%%%%%%%%%%%
%%%%%%%%%%%%%%%%%%%%%%%%%%%%%%%%%%%%%%%%%%%%%%%%%%%%%%%%%%%%%%%%%%%%%%%%%%%%%%%%%%%%%%%%%%%%
\section{Conclusion}
\label{sec: conclusion}

In this article we   presented a model for computing the evolution of the dust grains in a protostellar collapse. On the one hand, we   showed that the dynamical evolution of the gas is crucial for the evolution of the carbonaceous grains, notably because the chemisputtering by $\mathrm{H}$ atoms has a range of temperatures where the characteristic time has the same order of magnitude as the free-fall time. On the other hand, the silicate and aluminum oxide grains sublimate at a relatively fixed temperature, but the evolution of their size distribution is crucial in order to compute the non-ideal MHD resistivities and the opacities. We  also showed that the vaporization of the grains in complex trajectories with decreasing temperature can have a significant impact on the non-ideal MHD resistivities and the opacities of the medium, which are not adequately captured by a simple temperature-dependent vaporization model. Additionally, we   discussed the limitations of our model and the potential applications of our model to protoplanetary disks. Despite these limitations, the dynamical approach gives results that are quite different from the current one used in numerical simulations, notably in the case of fast temperature and density variation. This suggests that in the context of protostellar collapse or of dust grain ejection from outflows, the dynamical approach is necessary.

A potential follow-up to this work could be to include the chemical evolution of the gas phase, to take into account the nucleation of new grains, and to apply the model to protoplanetary disks. In this way, we could implement the vaporization of the grains in hydrodynamical codes to have a more realistic model of dust evolution in the protostellar collapse. The works of~\citet{lombartGeneralNonlinearFragmentation2024a} on grain coagulation and fragmentation could be extended to take into account the chemisputtering of the grains, and be coupled with various hydrodynamical codes, such as RAMSES~\citep{teyssierCosmologicalHydrodynamicsAdaptive2002} and IDEFIX~\citep{lesurIDEFIXVersatilePerformanceportable2023}.

\begin{acknowledgements}
   We thank the anonymous referee for their useful comments that have improved the quality of this paper. This work has received funding from the French Agence Nationale de la Recherche (ANR) through the projects PROMETHEE (ANR-22-CE31-0020), and DISKBUILD (ANR-20-CE490006). We thank Asmita Bhandare for providing us with the data of the collapse simulation from their~\citet{bhandareMixingEasyNew2024a} paper. We finally thank Sébastien Charnoz for the enriching discussions we had.
\end{acknowledgements}

\bibliographystyle{aa} % style aa.bst
\bibliography{DustPaper} % your references

\begin{appendix}
\section{Thermochemical data}
\label{app: thermochemical_data}

We list here the values of all the parameter needed to compute the vaporization of the grains. To compute the Gibbs energy of formation $\indrm{\Delta}{f}G$ of the different chemical species involved in the model, we use the standard enthalpy of formation $\indrm{\Delta}{f}^0H$ and the standard entropy $S^0$ of the species. The Gibbs energy of formation at a given temperature $T$ is then given by
\begin{equation}
   \indrm{\Delta}{f}G = \indrm{\Delta}{f}^0H - T S^0.
\end{equation}
The data for the species involved in reactions~\eqref{eq: free_evaporation_forsterite} to~\eqref{eq: free_evaporation_aluminium_oxide} are shown in Table~\ref{tab: thermochemical_data}.
The values for the sticking coefficients $\alpha$ used in the model are also shown in Table~\ref{tab: alpha_data} for information, but their precise values have little impact on the model results as long as the order of magnitude chosen is of the good order (here, between~0.1-0.3).
{ Finally, for the other gas species that are not directly involved in the vaporization process, but which still appear in the gas equilibrium computation by FASTCHEM2~\citep{stockFASTCHEM2Improved2022}, we use the data table available (file \texttt{logK.dat}) in the FASTCHEM GitHub repository\footnote{\href{https://github.com/NewStrangeWorlds/FastChem}{https://github.com/NewStrangeWorlds/FastChem}}, which is build using mainly the data from~\citet{chaseNISTJANAFThermochemicalTables1998}.}
\begin{table}[h]
   \centering
   \begin{tabular}{c|cc}
      \hline
      Species & $\indrm{\Delta}{f}^0H$ (kJ mol$^{-1}$) & $S^0$ (J K$^{-1}$ mol$^{-1}$) \\
      \hline
      $\mathrm{Al_{(g)}}$ & 329.699 & 164.553 \\
      $\mathrm{AlO_{(g)}}$ & 66.944 & 218.386 \\
      $\mathrm{Al_2O_{(g)}}$ & -145.185 & 252.332 \\
      $\mathrm{Al_2O_{3,(s)}}$ & -1675.692  & 50.950 \\
      $\mathrm{H_{2, (g)}}$ & 0 & 130.680 \\
      $\mathrm{H_2O_{\mathrm{(g)}}}$ & -241.826 & 188.834 \\
      $\mathrm{Mg_{(g)}}$ & 147.100 & 148.648  \\
      $\mathrm{Mg_2SiO_{4,(s)}}$ & -2176.935 & 95.140 \\
      $\mathrm{O_{2, (g)}}$ & 0 & 205.147 \\
      $\mathrm{SiO_{(g)}}$ & -100.416 & 211.579 \\
      \hline

   \end{tabular}
   \caption{Thermochemical data for the species involved in reactions~\eqref{eq: free_evaporation_forsterite} to~\eqref{eq: free_evaporation_aluminium_oxide}. The data are   from~\citet{chaseNISTJANAFThermochemicalTables1998}.}
   \label{tab: thermochemical_data}
\end{table}

\begin{table}[h]
   \centering
   \begin{tabular}{c|c}
      \hline
      Reaction & $\alpha$ \\
      \hline
      \eqref{eq: free_vaporization_carbon} & 0.3 \\
      \eqref{eq: free_evaporation_forsterite}-\eqref{eq: chemisputtering_forsterite} & 0.1 \\
      \eqref{eq: chemisputtering_aluminium_oxide_1}-\eqref{eq: free_evaporation_aluminium_oxide} & 0.25 \\
      \hline

   \end{tabular}
   \caption{Sticking coefficients $\alpha$ used in the model for the different reactions.}
   \label{tab: alpha_data}
\end{table}

\end{appendix}

\end{document}